\documentclass[useAMS,usenatbib]{mn2e}
\usepackage{times}
\usepackage{graphicx, latexsym, amssymb, amscd, psfrag}
\usepackage{epsfig}
\usepackage{color}
\usepackage{morefloats,rotating,float}
\title[GAMA: Nature of the dwarf galaxy population]{Galaxy And Mass Assembly (GAMA): The unimodal nature of the
 dwarf galaxy population }
\author[Mahajan et al.]
{\parbox{\textwidth}{Smriti Mahajan,$^{1,2}$\thanks{E-mail: \texttt{smritimahajan@iisermohali.ac.in}}
 Michael J. Drinkwater$^{2}$,  S. Driver$^{3,4}$, Lee S. Kelvin$^{5}$, A. M. Hopkins$^{6}$, I. Baldry$^{7}$,
 S. Phillipps$^{8}$, J. Bland-Hawthorn$^{9}$, S. Brough$^{6}$, J. Loveday$^{10}$,
 Samantha J. Penny$^{11}$, A.S.G. Robotham$^{3}$}\vspace{0.4cm}\\ 
 \parbox{\textwidth}{$^{1}$Indian Institute for Science Education and Research Mohali- IISERM, Knowledge City, Manauli, 140306,
 Punjab, India\\ 
 $^{2}$School of Mathematics and Physics, University of Queensland, Brisbane, QLD 4072, Australia\\
 $^{3}$International Centre for Radio Astronomy Research (ICRAR), 
University of Western Australia, Crawley, WA 6009, Australia\\
 $^{4}$Scottish Universities' Physics Alliance (SUPA), School of Physics and Astronomy,
  University of St Andrews, North Haugh, St Andrews, KY16 9SS, UK \\
 $^{5}$Institut f\"or Astro- und Teilchenphysik, Universit\"at Innsbruck, Technikerstrabe 25, 6020 Innsbruck, Austria\\
 $^{6}$Australian Astronomical Observatory, PO Box 915, North Ryde, NSW 1670, Australia\\
 $^{7}$Astrophysics Research Institute, Liverpool John Moores University, IC2, Liverpool Science Park, 146 Brownlow Hill, Liverpool, L3 5RF\\
 $^{8}$Astrophysics Group, School of Physics, University of Bristol, Bristol BS8 1TL, UK\\
 $^{9}$Sydney Institute for Astronomy, School of Physics A28, University of Sydney, NSW 2006, Australia\\
 $^{10}$Astronomy Centre, University of Sussex, Falmer, Brighton BN1 9QH, UK\\
 $^{11}$School of Physics, Monash University, Clayton, Victoria 3800, Australia} }

\newcommand{\s}{starburst}
\def\eg{{e.g. }}
\def\g{{GAMA }}
\def\s{{{\sc sigma }}}

\def\nr{{$NUV-r$}}
\def\mue{{$\langle \mu \rangle_e$}}
\def\re{{$R_{eff}$}}

\definecolor{grey}{rgb}{0.5,0.6,0.7}

\begin{document}

\date{}

\pagerange{\pageref{firstpage}--\pageref{lastpage}} \pubyear{2013}
\maketitle

\label{firstpage}
\begin{abstract}
In this paper we aim to (i) test the number of statistically
distinct classes required to classify the local galaxy population, and, (ii) identify the differences in the physical
and star formation properties of visually-distinct galaxies. To accomplish this, we analyse the structural parameters (effective radius (\re), effective surface brightness within \re~(\mue), central surface brightness ($\mu_0$), and S\'ersic index ($n$)), obtained by fitting the light profile of 432 galaxies ($0.002<z\leq0.02$; {\it Viking} $Z$-band), and their spectral energy distribution using multi-band photometry in 18 broadbands to obtain the stellar mass ($M^*$), the star
 formation rate (SFR), the specific SFR (sSFR) and  the dust mass ($M_{dust}$), respectively. 

We show that visually distinct, star-forming dwarf galaxies (irregulars, blue spheroids and low surface brightness galaxies) form a unimodal population in a parameter space mapped by \mue, $\mu_0$, $n$, \re, SFR, sSFR, $M^*$, $M_{dust}$ and ($g-i$). 
The SFR and sSFR distribution of passively evolving (dwarf) ellipticals on the other hand, statistically distinguish them from other galaxies with similar luminosity, while the giant galaxies clearly segregate into star-forming spirals and passive lenticulars. We therefore suggest that the morphology classification scheme(s) used in literature for dwarf galaxies only reflect the observational differences based on luminosity and surface brightness among the apparent distinct classes, rather than any physical differences between them.

\end{abstract}


\section{Introduction}
\label{paperintro}

 The visual appearance of a galaxy is the manifestation of its star formation history, stellar content and, dynamic and kinematic 
 properties. As a result, visual morphology of a galaxy can not only provide important clues to its formation mechanisms, but the 
 longevity of specific features such as the spiral structure and bars could potentially be used to understand secular evolution.
 
 It is however a known fact that galaxy properties are modulated by their environment, such that red, bulge-dominated,
 passively evolving galaxies are mostly found in the dense cores of galaxy clusters and groups, while blue, star-forming, disky, galaxies
 appear more frequently in less dense environments \citep[e.g.][]{dressler80}. But can a galaxy evolve from being disky to spheroidal 
 away from the harsh environment of clusters and groups? 
 To learn about the mechanisms responsible for converting one galaxy population into another, and, any evolutionary links
 between visually distinct galaxies, it is important to get a complete description of galaxies which are evolving
 independent of the impact of dense environment.
  
 Historically, galaxies were distinguished by visual classification \citep{hubble26}, supported by quantitative description of
 their stellar light profile \citep{reynolds20}. But since Hubble's tuning fork diagram, we have come a long way in classifying
 galaxies based on spectroscopic or photometric data obtained in multiple wavebands. After morphology, the most popular
 classification scheme for galaxies is based on luminosity. Most studies based on wide-angle survey data such as those obtained by the
 Sloan Digital Sky Survey (SDSS), define galaxies fainter than $M_r\sim-18$\,mag as dwarfs.
 One of the key characteristics of low-luminosity galaxies however, is their low surface brightness which is correlated with
 their luminosity albeit with considerable scatter \citep{caldwell83,binggeli84}. However, in some cases like the giant low
 surface brightness (LSB) galaxy Malin I \citep{bothun87}, and galaxies having extended ultraviolet disks \citep[e.g.][]{thilker07},
 the need to distinguish between low-surface brightness and low-luminosity galaxies might be more critical than immediately intuitive. 
 
 The literature on low-luminosity, low surface brightness galaxies is
 further complicated by incorporation of extensive jargon over the decades. Although dwarf ellipticals (dEs), (dwarf) irregulars ((d)Irr),
 dwarf spheroidals (dSph), dwarf spirals or lenticulars (dS, dS0), blue compact dwarfs (BCDs), ultra-compact dwarfs (UCDs),
 and little blue spheroids (LBS) all have their glorious presence in the extensive literature, it is often not obvious whether
 the difference between these subpopulations of galaxies extend beyond their morphology, or if one class is related to another
 or their giant counterparts. 
  
 While dwarf ellipticals, which are predominantly found in rich galaxy clusters, have received considerable attention in the context 
 of the dichotomy between dEs and giant ellipticals \citep[\eg][and references therein]{kormendy85,graham03,kormendy12,graham13},
 and, more recently for structure `hidden' under the smooth radial profiles \citep{janz12,janz13}, other classes of low-luminosity 
 galaxies have not been studied much beyond the local volume ($\lesssim10$\,Mpc). This is partly due to the lack of data and partly due to the 
 low volume density of these galaxies, which are hard to observe and easily missed from observations focussing on interiors of galaxy clusters.
 Within the local volume however, dwarf galaxies, especially the dwarf irregulars have been well explored at the optical
 \citep[e.g.][]{sharina08,herrmann13} as well as the ultraviolet wavelengths \citep{hunter10}.
  It is however only with the advent of wide-angle sky surveys in the last decade or so that it is finally possible to measure a sufficient volume 
  of the local universe to sample large numbers of low-luminosity stellar systems outside the extreme environments of clusters.

%

  One such programme is the Galaxy and Mass Assembly survey \citep[GAMA;][]{driver11}, which now provides a unique
 opportunity to analyse the properties of low-luminosity, low surface brightness galaxies because of its very high volume (almost 20 000 Mpc$^3$) at low redshifts. The GAMA survey regions were not
 selected on the basis of environment.
  With photometric and spectroscopic information for nearby galaxies in hand, in this project we aim to:
 \begin{itemize}
 \item characterise the galaxies, especially dwarfs, residing away from clusters, 
 \item compare the intrinsic and physical properties of visually distinct galaxies with an aim to find evolutionary links between
 them, and
 \item determine the minimum number of classes which are required to separate galaxies, especially dwarfs,
  into statistically distinguishable populations.
 \end{itemize}
 At this point we emphasise that our unique sample (8-87 Mpc) helps in bridging the gap between studies of
 galaxies in the local volume 
 \citep[$\lesssim10$\,Mpc;][]{sharina08,mcconnachie12} and statistical analyses of galaxies in rich clusters such
 as Virgo ($16.5$\,Mpc) and Coma ($\sim100$\,Mpc).
 
 We characterise the dataset used in this paper and various selection biases that may affect our analysis in the following section. In
 \S\ref{structure} we explain the methods we used for classifying our sample: (i) the visual classification scheme, (ii) the light
 profile fitting algorithm 
  which is used to quantify the morphology of galaxies through structural parameters, and, (iii) the spectral energy distribution fitting
  procedure used to estimate the star formation and dust properties of galaxies. We discuss the relation between various structural
  parameters obtained from fitting the light profile in \S\ref{photo}, and discuss the star formation and dust properties of our sample in \S\ref{sf}.
 Various implications of our analysis on the stellar mass function of galaxies and the unimodal nature of galaxies are discussed in \S\ref{discussion}, along with the limitations of our analysis. We conclude with a summary
 of our findings in \S\ref{summary}. 

 All distances, magnitudes and masses in this paper are calculated under the assumption of a $\Lambda$CDM concordance
 cosmological model with $H_0=70$\,km s$^{-1}$ Mpc$^{-1}$, $\Omega_\Lambda=0.7$ and $\Omega_m=0.3$. 
  
 \section{Data and sample selection}
 \label{data}
 
 \subsection{GAMA spectroscopic data}
 
The Galaxy and Mass Assembly (GAMA) survey is a combined spectroscopic and multi-wavelength 
programme that exploits various ground-based and space-borne observing facilities to study cosmology and
galaxy evolution. The \g spectroscopic campaign is based on the Sloan Digital Sky Survey  (SDSS; Data Release 7) imaging
complete to $r=19.8$. The survey, after completion of Phase II will have photometry in 20 wavebands,
from far ultraviolet to radio, and, spectroscopic redshifts for $\sim 300,000$ galaxies ($z\lesssim 0.25$) in three equatorial
 regions, each measuring $12\times5$ square degrees and two southern fields of similar size \citep{driver11}.
 
\begin{figure}
 \centering{
{\rotatebox{270}{\epsfig{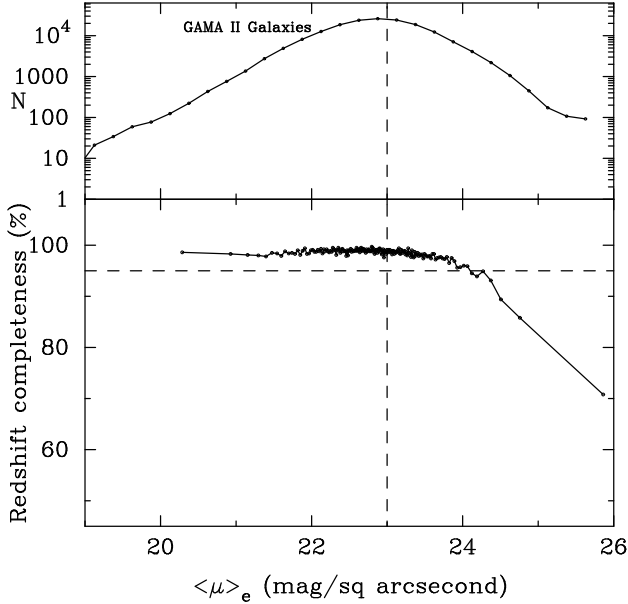}}}}
\caption{{\it (Top)} The number distribution of all \g II sources with $r\leq 19.8$ and reliable redshift. 
 {\it (Bottom)} Redshift success rate for all \g II galaxies as a function of surface brightness in bins of 1000 galaxies. 
 The horizontal and vertical dashed lines represent 95\% redshift completeness and surface brightness completeness
 limit of $\mu_{r}=23$ magnitude per square arcsecond adopted here \citep[also for SDSS; see][]{blanton05}, respectively. }
\label{sb}
\end{figure}

The surface brightness completeness limit for the \g survey is mainly driven by the SDSS imaging, and is important
 for understanding some of the selection biases in our sample. Fainter than this completeness limit the imaging data becomes
 unreliable. Since the targets for the spectroscopic campaign for \g are selected in the $r$-band, in Figure~\ref{sb}
 we show the number distribution of all the \g II sources with $r\leq 19.8$, SURVEY\_CLASS$>1$ and VIS\_CLASS$\leq1$ or
 $=255$ \citep{baldry10}. The VIS\_CLASS parameter signifies the likelihood of an object for being a target for spectroscopic follow-up.
Many of the low-redshift targets are classified as VIS\_CLASS$=3$, implying they are deblended part of a galaxy, and are
 excluded from our analysis. 

 The bottom panel of Figure~\ref{sb} shows the fraction of galaxies with reliable redshifts
 \citep[quality of redshift, $nQ\geq3$;][]{baldry10} in each bin of surface brightness, in this sample.
 As expected, the completeness fraction drops towards fainter surface brightness and goes below 95\%
 at $\mu_e>24.2$\,magnitude per square arcsecond. This is in agreement with the median 1$\sigma$ limit of 
 $\mu_e=24.60$\,magnitude per square arcsecond derived for \g I
 \citep[$r_{complete}=19.4$;][]{kelvin12}. Since we are mainly dealing with low-luminosity and low-surface brightness galaxies in
 this paper, we adopt a conservative surface brightness completeness limit of $\mu_e=23.0$\,magnitude per square arcsecond, consistent
 within the limit advocated in \citet{blanton05}.
  
 The sample used in this work is selected from the 
 \g II catalogue such that $nQ\geq3$ and $0.002<z_{TONRY}\leq0.02$ \citep[see][for details on the local flow correction]{baldry12}.
  This redshift range is optimally chosen to exclude galactic stars and enable visual classification of low-luminosity and low surface brightness 
 galaxies in the shallow imaging data from the SDSS. All images were visually inspected for any remaining 
 artefacts, stars and duplicates resulting in the final catalogue of 432 galaxies (including those fainter
 than $\mu_e=23.0$\,magnitude per square arcsecond) which is used throughout this paper unless stated otherwise.

 All magnitudes used in this work are corrected for galactic extinction and K-corrected to $z=0$ using kcorr\_z00v03 \citep{loveday12}
 and GalacticExtinctionv02 data management units for \g II. 
 While \g II is 99\% complete for galaxies down to $r=19.8$, Figure~\ref{sb} shows that the surface brightness incompleteness 
 starts becoming significant at \mue$>24.2$\,magnitude per square arcsecond. This limit thus restricts the overall angular size of
 the detected galaxies such that $\mu_{limit}=m+2.5log(2\pi R_{limit}^{2})$, where $\mu_{limit}$, $m$ and
 $R_{limit}$ are the limiting effective surface brightness, magnitude and limiting size for the sample, respectively \citep[see][for a detailed
 derivation of the formula]{graham05}. 
 
 \subsection{The {\it Viking} data}
 
 The VISTA Kilo-degree Infrared Galaxy survey ({\it Viking})
 will cover 1,500 square degrees of sky in five broadband filters ($Z,Y,J,H,K_s$) with the 4.1 metre Visible and Infrared Survey Telescope for
 Astronomy (VISTA), located at the Paranal Observatory in Chile. The {\it Viking} data has $2\times$ better resolution ($0.6^{\prime\prime}$), 
 and depth approximately two magnitudes deeper than the SDSS. 

 Since the \g parent sample was selected in the $r$-band, while here we choose to present the analysis in the {\it Viking} $Z$-band,
 we need to evaluate the limiting values for all parameters in the $Z$-band. 
 The difference between the SDSS $r$-band and the {\it Viking} $Z$-band magnitudes for our sample galaxies is 
 in the range $-1.69<m_r-m_Z\leq1.96$, with a mean difference of 0.31 magnitude and standard deviation of 0.36 magnitude, respectively.
 For $\sim95\%$ of the galaxies with data in both wavebands, the mean difference in the effective surface brightness within the
 effective radius is 0.35 magnitude per square arcsecond. Based on the above mentioned differences, in the following we assume a
 limiting effective surface brightness of $\mu_{limit}=22.65$\,magnitude per square arcsecond. The limiting magnitude
 $m_{limit}=19.5$\,mag at our $z_{max}=0.02$ in the $Z$-band translates to $M_{Zlimit}=-15$. 

 \subsection{Environment of galaxies}

 As we discuss in Section~\ref{paperintro}, most previous studies of low-luminosity galaxies focus on the dense environment of
 rich galaxy clusters. There are however, no massive clusters or groups in any of the three equatorial \g regions at $z\leq0.02$ as
demonstrated by the redshift distribution of \g galaxies in figure 5 of \citet{baldry12}.
In order to quantify environment for our sample, we also made use of 
 the GAMA group catalogue \citep{robotham11}. However, due to the high lower redshift limit of this catalogue, 
 the clustering information based on the friends-of-friends method is only available for a subsample of 263 of our galaxies 
 at $z>0.01$.
 Of these 263, 11 galaxies are found in groups with multiplicity $\geq3$, and another 25 are in pairs. 

 Together, these properties demonstrate that the galaxies in our sample avoid the very dense cluster environments 
 typical of most previous dwarf galaxy samples.
 This does not however imply that they are entirely isolated, but do satisfy our aim
 to determine the properties of galaxies evolving away from the influence of dense environments.

 \section{Methodology}
 \label{structure}
 We adopt three independent methods for the classification of galaxies: visual classification based on inspecting
 the five-colour SDSS images, quantitative structural analysis based on the modelling of the light profile of galaxies, and star formation
 properties obtained by fitting the spectral energy distribution of galaxies. We describe each of these methods below.
 
 \subsection{Visual classification of galaxy morphology} 
 
 \begin{figure*}
\centering{
{\rotatebox{0}{\epsfig{file=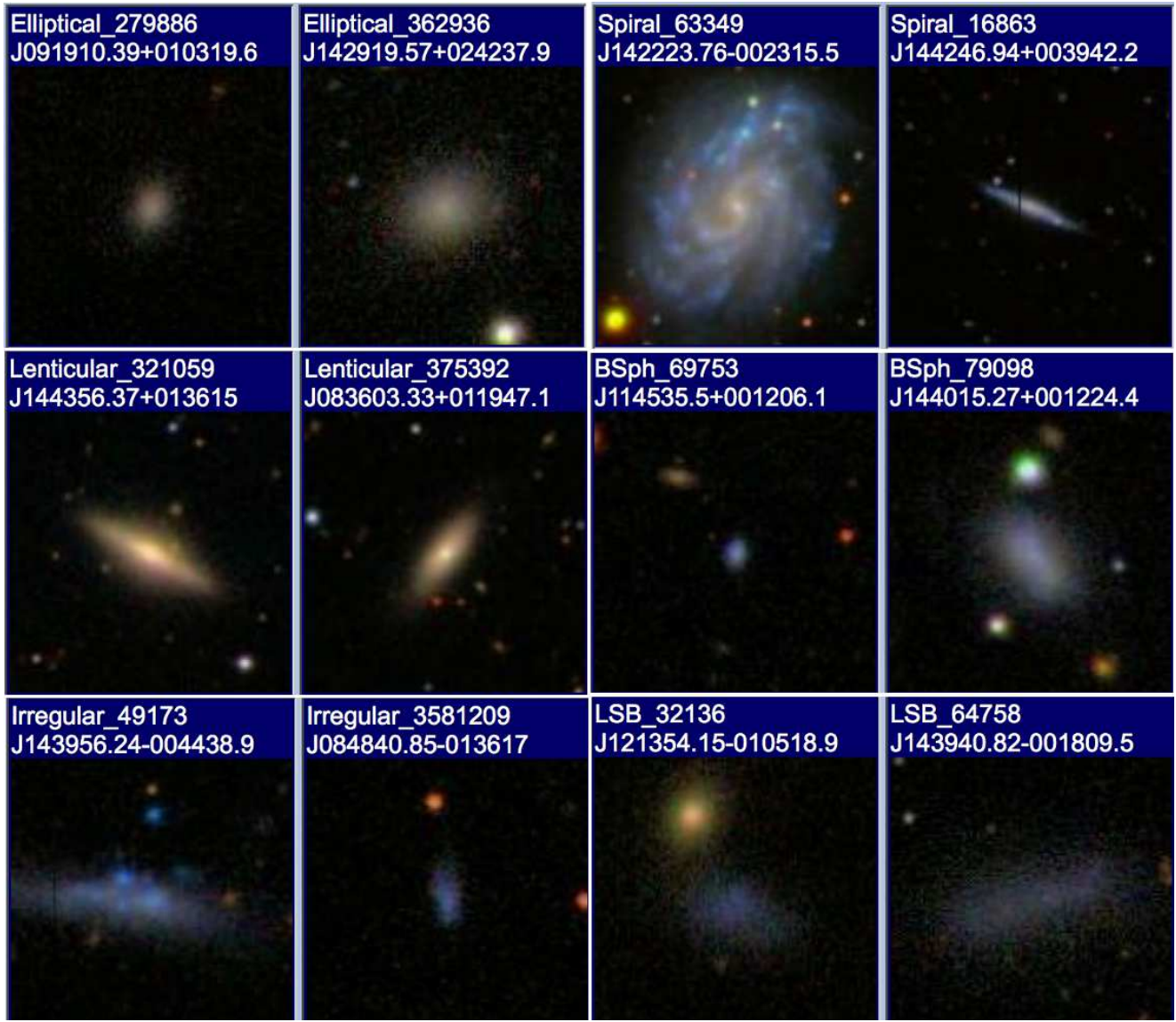,width=17.5cm}}}}
\caption{Montage of representative galaxies from each of the six visually identified morphology classes. The \g II
 ID and class are shown at the top of each image followed by its objID from SDSS(DR7). }
\label{irr}
\end{figure*}

 The visual appearance of galaxies has historically been used to understand their evolutionary sequence and formation mechanisms
 by separating them into classes. The Hubble tuning fork diagram \citep{hubble26}, for instance is in use to the present day.
 The Hubble sequence, which was designed to be an evolutionary sequence in morphological types, can 
 also be interpreted as a sequence in gas content, mass, and bar
 structure of galaxies, all of which are in a symbiotic relationship with their global star formation rate (SFR) and dynamic environment. 
   
 One of our key motivations for this project is to test if galaxies that can visually be classified into separate classes, are statistically
 distinguishable in other properties, indicative of an evolutionary link between them. To accomplish this, Mahajan 
 visually classified the five colour SDSS images of all galaxies ($0.002\!\leq\!z\!\leq\!0.02$) five times, classifying them into six
 categories as detailed below. In this redshift range the physical resolution of the SDSS images is around $0.05-0.48$kpc. 
 The classification scheme was designed without any presumptions about how many categories were required. In other words, a new
 category was created based on the imaging data if there appeared to be galaxies which 
 shared common traits in their appearance, but did not fit into any of the existing classes. Mahajan, Driver and Drinkwater also visually
 classified the three-colour images in the $giH$ wavebands. On average, the various classifiers agreed 80\% of the time with the
 low-surface brightness galaxies and blue spheroids with red centres appearing as the most doubtful cases.
 Amongst different types, different classifiers disagreed with themselves and with each other for galaxies mostly
 classified as irregulars; a non-negligible fraction of the irregular galaxies were equally likely to be classified as an LSB
 or a blue spheroid. 

 The visual classification scheme was designed to categorise all galaxies using their broadband colour and morphology. 
 Where possible however, broadband colour was not accounted for in order to simplify the morphology classification scheme (Figure~\ref{irr}). 
 The different classes and total number of galaxies in each class are:
\begin{itemize}
\item {\it Elliptical (E; 21):} Galaxies which are morphologically elliptical in shape. They are mostly red in colour.  
 \item {\it Spirals (Sp; 47):} Galaxies showing well-defined spiral arms or clearly identifiable edge-on disks. These galaxies
 often show conspicuous signs of ongoing star formation, such as HII regions, and stellar associations forming spiral arms.
 \item{\it Lenticulars (L; 26):} Red, disk galaxies with a resolved nucleus. These galaxies are mostly big and bright, occasionally showing
 signs of some ongoing star formation in rings around nucleus, or low surface brightness disks without spiral arms.
 \item {\it Blue spheroidals (BSph; 73):} Colour plays a key role in successfully identifying these galaxies. They are very blue
 and generally compact spheroids, morphologically similar to small elliptical galaxies or bulges of spiral galaxies.
 \item{\it Low surface brightness (LSB; 69) galaxies:} These extended objects show very poor contrast with the
 background in the five-band SDSS imaging. We note that a substantial fraction of these galaxies may have been
 misclassified due to the very shallow imaging data used here. Many of these galaxies may also be classified as
 irregular, and as we will show below, these two classes overlap in most of the parameter space explored here.  
 \item {\it Irregulars (Irr; 196):} All confirmed extended sources that do not belong to any of the above categories.
\end{itemize}
Some representative examples from each of the six classes are shown in Figure~\ref{irr}.

 We did not use luminosity in our classifications and avoided labelling galaxies as dwarfs or giants based on appearance alone 
 because the angular size of galaxies is a function of redshift. However, the resulting classifications were a strong function
 of luminosity with the last three classes listed above (BSph, LSB and Irr) dominating the low luminosity galaxies (as shown
 in Figure~\ref{m-r}). The mean fraction of these galaxies in the whole sample is 78 per cent, with the fraction increasing from zero
 at $M_Z\sim -20$ to 100 per cent at $M_Z\sim-16$; it reaches the mean fraction at a luminosity of $M_Z=-18.5$. This is equivalent to the
 $M_r\sim-18$\,mag limit for dwarf galaxies discussed above. We therefore refer to galaxies with $M_Z>-18.5$ collectively as dwarfs
 in order to make the following discussions more concise, unless stated otherwise. 
 
\subsection{Structural investigation of galaxy morphology with model analysis {\sc (sigma)}}
\label{sigma}

 \begin{figure*}
\centering{
{\rotatebox{0}{\epsfig{file=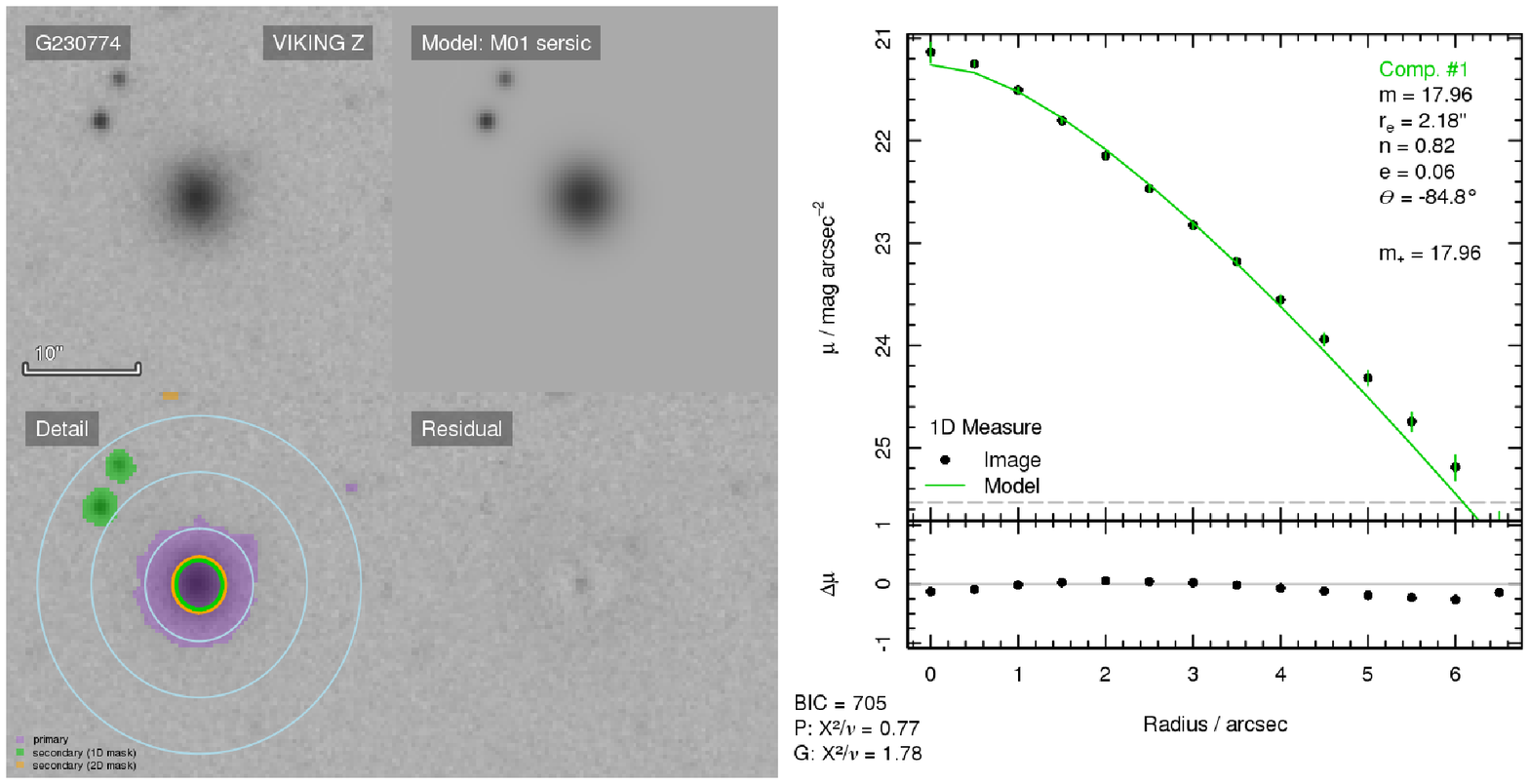,width=8.5cm}}}
{\rotatebox{0}{\epsfig{file=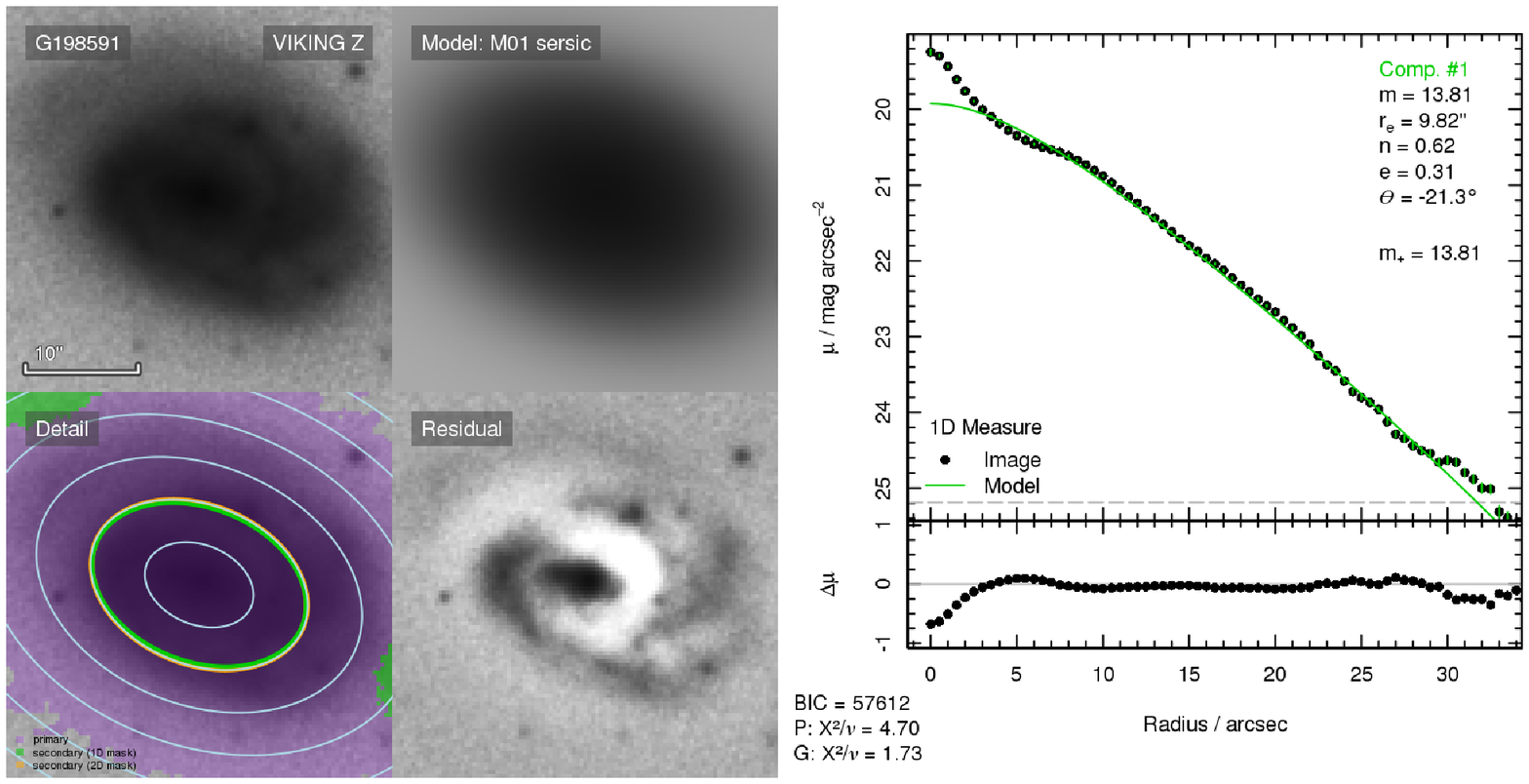,width=8.5cm}}}
{\rotatebox{0}{\epsfig{file=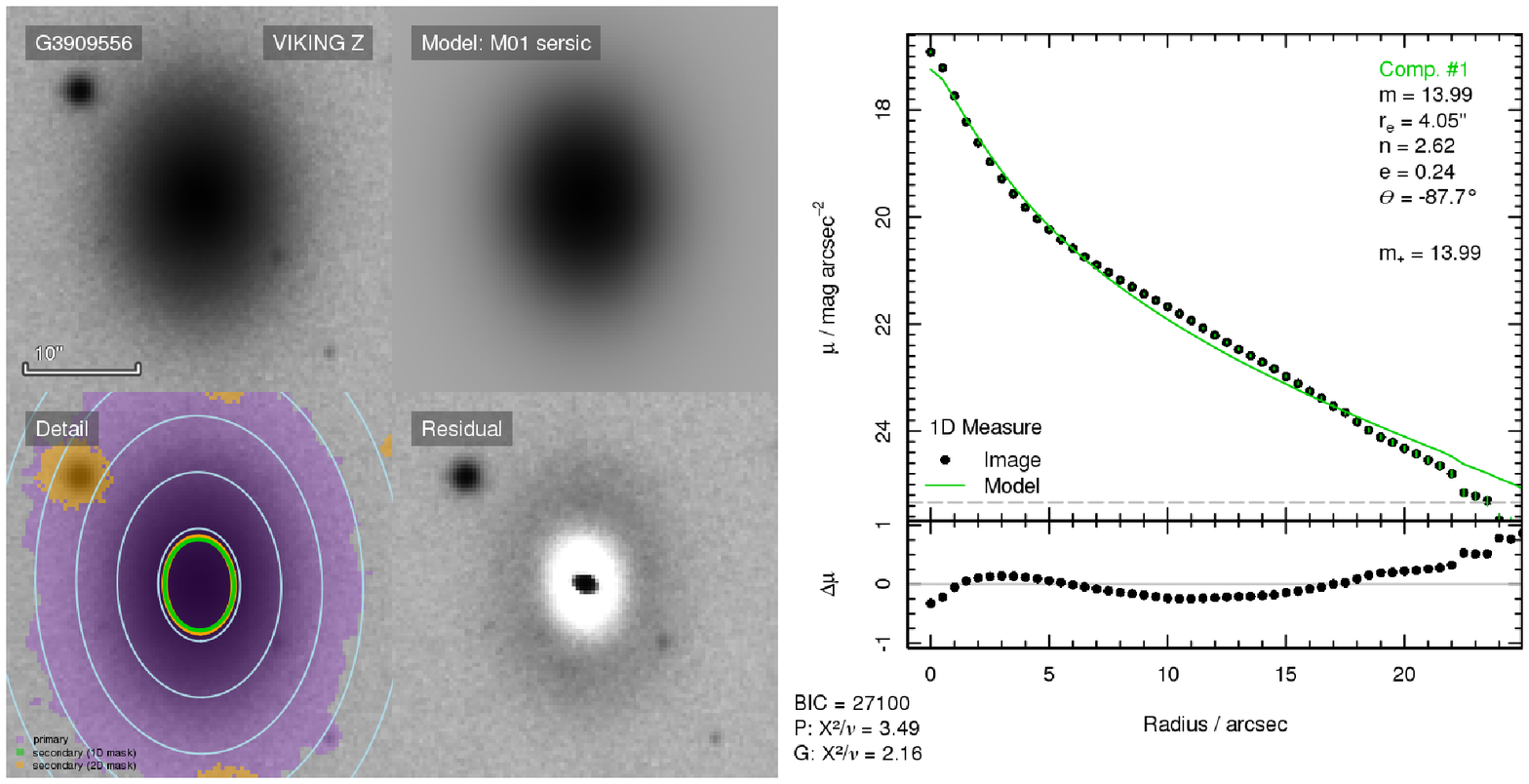,width=8.5cm}}}
{\rotatebox{0}{\epsfig{file=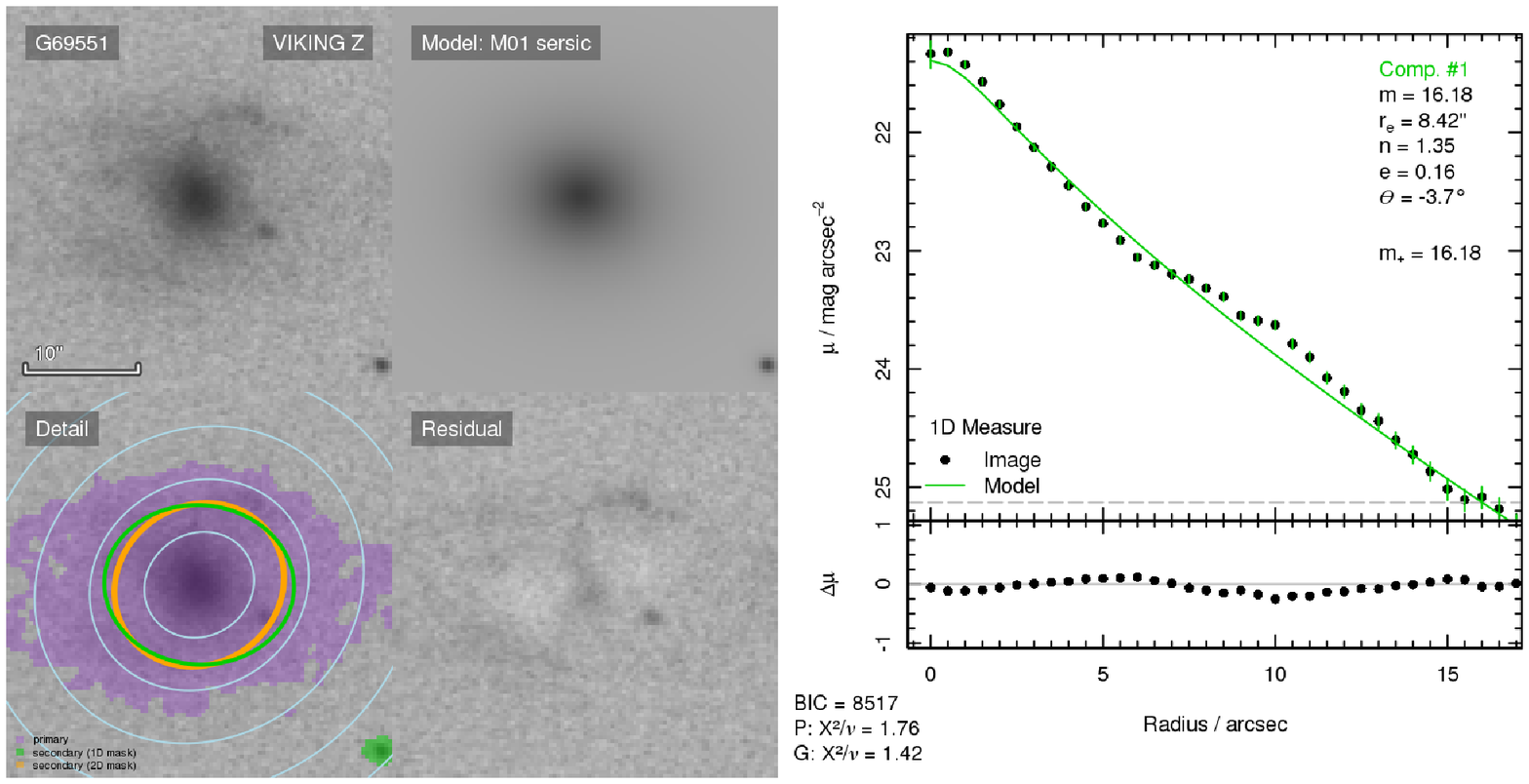,width=8.5cm}}}
{\rotatebox{0}{\epsfig{file=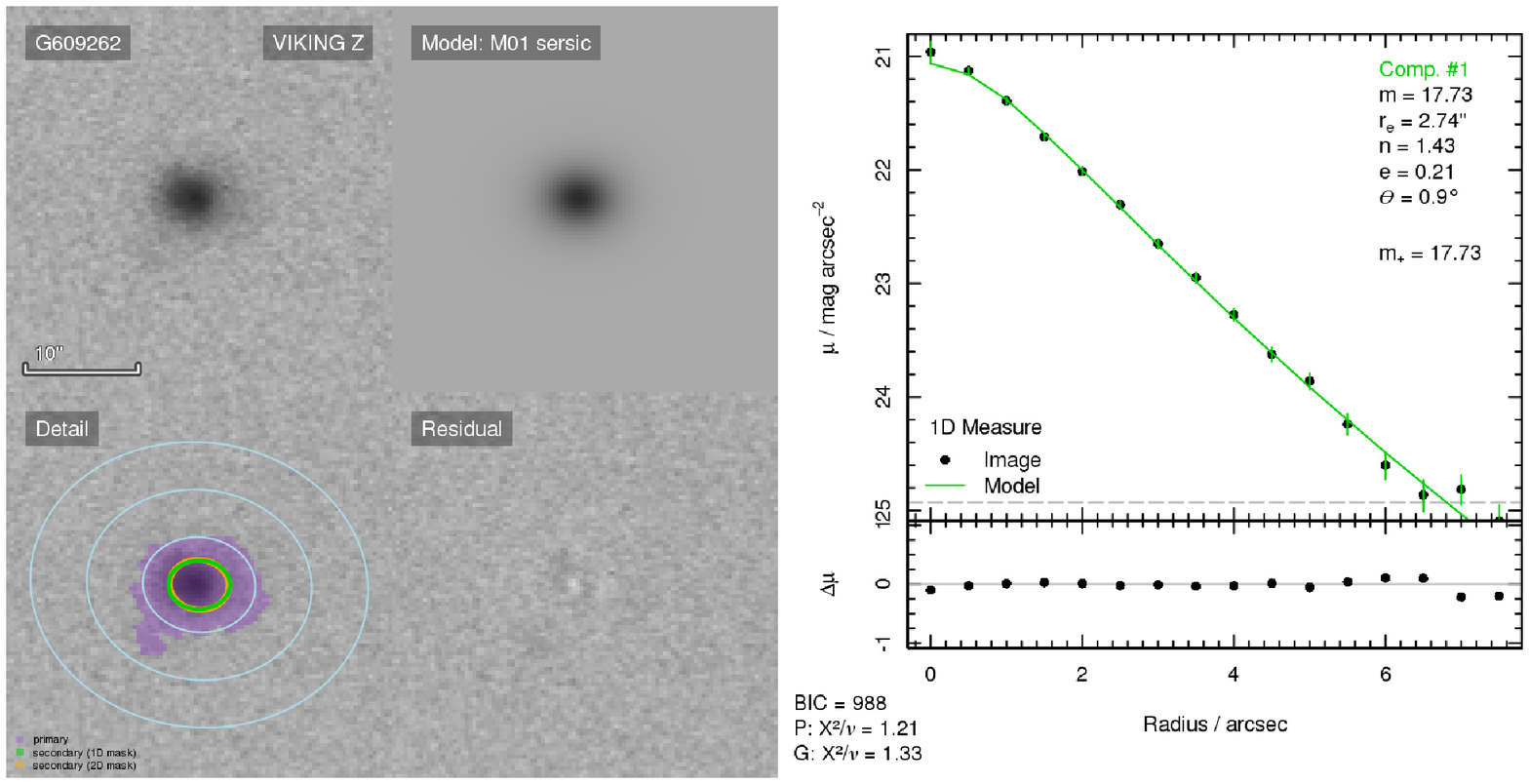,width=8.5cm}}}
{\rotatebox{0}{\epsfig{file=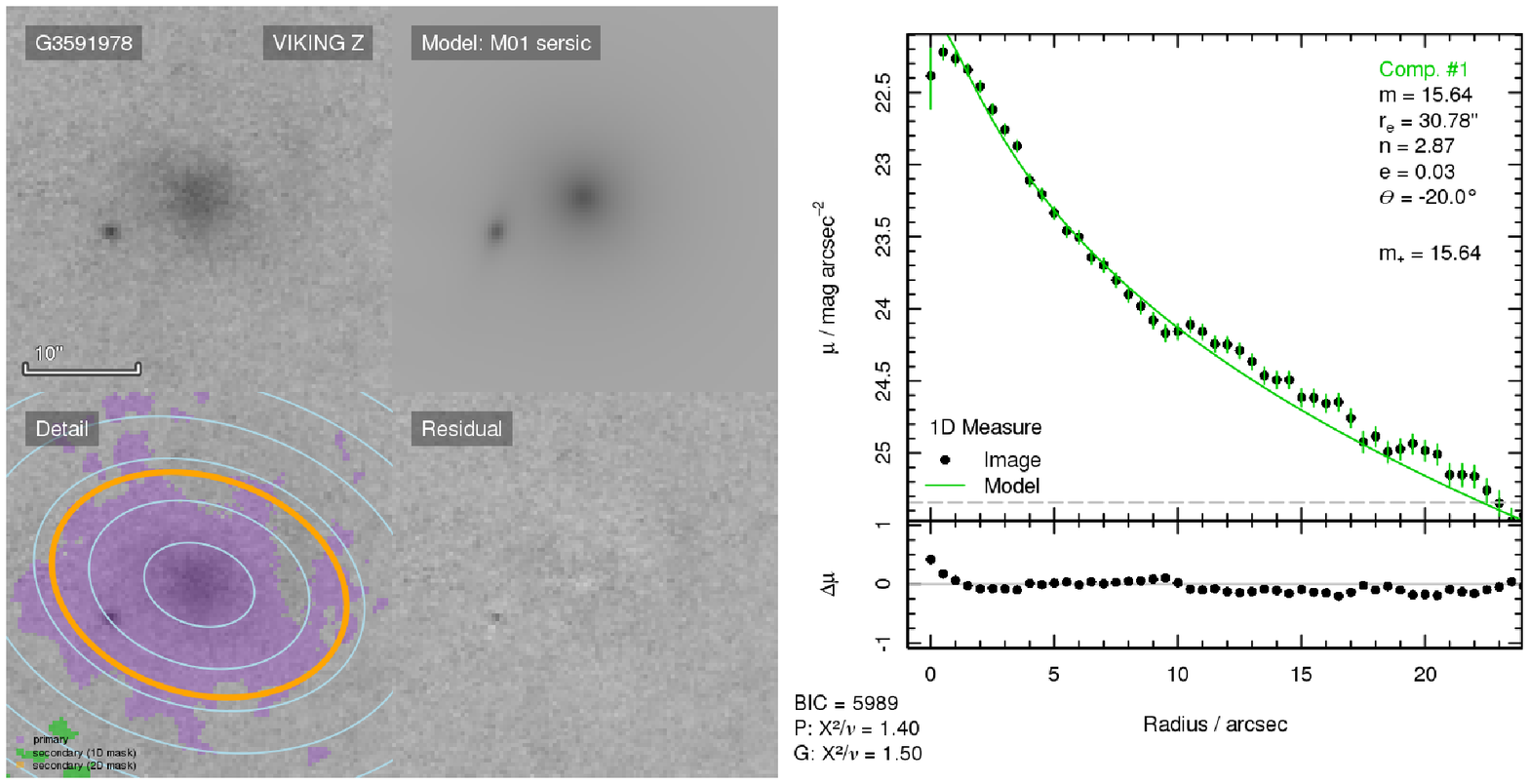,width=8.5cm}}}}
\caption{The surface brightness fits in the {VIKING} $Z$-band for some of the galaxies in our sample.
 {\it (left to right from top:)} An elliptical, spiral, lenticular, irregular, blue spheroid and low surface brightness galaxy, respectively. 
 Each panel shows {\it (clockwise from top left:)} the original Viking $Z$-band image, S\'ersic model, 1D light profile (with residuals:
 image-model at the bottom), residual image and, ellipses centred on the primary galaxy used for estimating the light profile along with
 masked objects, respectively. }
\label{sigma}
\end{figure*}

 \s \citep{kelvin12} employs a range of image analysis software and logical filters to perform structural analysis on an input
 catalogue of galaxies. At the heart of \s lies {\sc source extractor} \citep{bertin96}, {\sc psf extractor}
 \citep[{\sc psfex};][]{bertin11} and {\sc galfit 3} \citep{peng10}, which are aided by additional packages in the fitting
 process. Detailed explanations for various packages and their respective roles in the fitting process are given by 
 \citet{kelvin12}. \s only requires the image and the position of the primary galaxy therein as an input. It then outputs
 magnitude, fitted sky background and the fit parameters described below.
 
 {\sc galfit} fits a single S\'ersic function to each primary galaxy with seven free parameters: object centres $x_0$
 and $y_0$, total integrated magnitude $m_{tot}$, effective radius along the semi-major axis \re, S\'ersic
 index $n$, ellipticity and position angle. Secondary objects in the image are modelled by either a single
 S\'ersic function or a scaled point spread function (PSF) for stars, as appropriate. 
 The PSF comprises three free parameters $x_0$, $y_0$
 and $m_{tot}$. See \citet{peng10} for more information on the fitting process. A single S\'ersic function which describes
 the light profile of a galaxy as a function of its radius is given by the S\'ersic equation
 \begin{equation}
 I(r)=I_{e} \rm exp\left[-b_{n}\left(\left(\frac{r}{R_{eff}}\right)^{1/n}-1\right)\right],
 \end{equation}
where $I(r)$ is the intensity at radius $r$, $I_e$ is the intensity at the effective radius \re, the approximate radius containing
 half of the total light, and $n$ is the 
S\'ersic index which determines the shape of the light profile of a galaxy. The value of $b_n$ is a function of the S\'ersic
 index. A large number of profile shapes can be fitted by varying $n$; $n=0.5$ gives a Gaussian profile, $n=1$
 an exponential profile suitable for galaxy discs, and $n=4$ a de Vaucouleurs profile mostly associated with spheroids
 such as giant elliptical galaxies \citep[see][for a detailed description of the S\'ersic model]{graham05}. The quality of fits
 can be judged from Figure~\ref{sigma}, which shows a typical example of fitted light profile for each morphology
 class discussed above. Although $>91\%$ of our sample galaxies are fitted well by {\sc sigma}, the single component S\'ersic model
 does not fit the centre of the nucleated larger galaxies, spirals and lenticulars, well.  
 The number of galaxies in each morphology class whose light profile could not be adequately fitted by \s are presented
 in Table~\ref{fit-stats}. We note that some of the galaxies may have more than one issue and hence are counted more than
 once for this table.  We discuss some limitations of the single component S\'ersic model and its impact on the presented
 analysis further in \S\ref{limit}.    

 It is however notable that only 2/21 elliptical galaxies in this \g sample may require a
 nuclear component, suggesting that in contrast to the (dwarf) elliptical galaxies in clusters
 \citep{thomas08}, most of the low-luminosity elliptical galaxies are not nucleated \citep[also see][]{oh2000}.  

  \begin{table*}
\caption{Success rate for {\sc sigma} and {\sc magphys} fitting for different morphological classes.}
\begin{tabular}{|l|c|c|c|c|c|c|}
\hline 
 Class $\rightarrow$           & Elliptical & Lenticular & Spiral & BSph & Irr  & LSB \\
 Number (Percentage) of galaxies in each class $\rightarrow$ &  21 (4.86) & 26 (6.01)   & 47 (10.88) &  73 (16.89) & 196 (45.37) & 69 (15.97) \\
Issues $\downarrow$ &     &        &    &     & &    \\ \hline \hline
{\sc sigma} fails to fit the centre of the galaxy        & 2           & 8             & 21       &            & 15        & 7 \\
{\sc sigma} fails to fit the outskirts of the galaxy      &             & 4             & 4          & 6         & 16      & 3  \\
{\sc sigma} fits a profile with $\Delta\mu>|0.2|$ in at least one bin     & 2         & 5              &  7        &   2        &  18     &  6  \\
{\sc sigma} fails to fit a surface brightness profile      &            &  1           &  8          &             &  13     &  5  \\
Companion object(s) interfering with the light profile fitting &  2        &   1          &            &  2            &  12     & 3  \\ 
{\sc magphys} fit not good  & 1      &  1           &   5           &              &  5        & 1  \\
{\sc magphys} fails to fit the SED &      &         &               &  1             &  1       &     \\
 \hline
\end{tabular}
\label{fit-stats}
\end{table*}
 
 No explicit constraints are placed on the parameters fitted by {\sc sigma}. However,
 {\sc galfit} has an internal limitation for S\'ersic index such that $0.05<n<20$, where the lower limit is a `soft' limit and the upper bound
 is a hard limit \citep{kelvin12}. 

\subsection{Multi-wavelength Analysis of Galaxy Physical Properties ({\sc magphys})}
\label{mag}

The spectral energy distribution of all the galaxies in our sample was fitted using the Multi-wavelength Analysis of Galaxy Physical Properties
 \citep[][{\sc magphys}]{dacunha08}.
The 18 band photometry was derived from cross-matching a number of distinct catalogues which include: the far ultraviolet ($FUV$) and
 near ultraviolet ($NUV$)
 {\it GALEX} data (GalexMainv02; best\_mag\_nuv, best\_magerr\_nuv, best\_mag\_fuv, best\_magerr\_fuv; see Seibert et al.~in prep);
 the aperture matched $u-K$ photometry (ApMatchedv04; Kron aperture matched photometry using SExtractor, see \citet{hill11} for details);
 publicly available WISE data \citep[][http://irsadist.ipac.caltech.edu/wise-allsky/; w(1,2,3,4)mpro and w(1,2,3,4)sigmpro]{wright10},
 and bespoke measurements from the Herschel-Atlas SPIRE observations (18BandPhotometryv01) using the $r$-band defined apertures
 convolved to the relevant SPIRE band point-spread function, and taking care to apportion flux from overlapping targets as described
 in Appendix A1 of \citet{bourne12}. A full description of the analysis of the assembly of these data will appear in Driver et al. (2014, in prep).

To overcome the background confusion level and the bright high redshift interlopers in the SPIRE bands, measurements were also made in
 apertures of identical area to the target galaxies but placed at random locations across the SPIRE data (Driver et al. 2014, in prep).
  The mean flux in the random apertures
 were then subtracted from the flux measured at the location of the target galaxy, and a flux error assigned based on the quadrature combination of the remaining measured target flux and the variance of the random measurements.

 The 18 band measurements were then converted from magnitudes to Janskys and fed into {\sc magphys} \citep{dacunha08}.
 {\sc magphys} then compares the data to an extensive library of stellar population and dust templates to derive fundamental measurements
 given flux measurements, the redshift and the known filter-set. For each galaxy, {\sc magphys} provides both best-fit values for
 derived parameters, and marginalised median and quartile values. Here we adopt the median and quartile values (which on occasion
 may lie significantly adrift from the best-fit values), for the following parameters: stellar mass $M^*$, dust mass $M_{dust}$, star formation
 rate and the specific star formation rate.

 \section{The structural parameters of morphologically distinct galaxies}
\label{photo}

In this section we analyse the relation between various structural properties of galaxies estimated by fitting their light profile in 
the {\it Viking} $Z$-band with a single component S\'ersic model.

 \begin{figure}
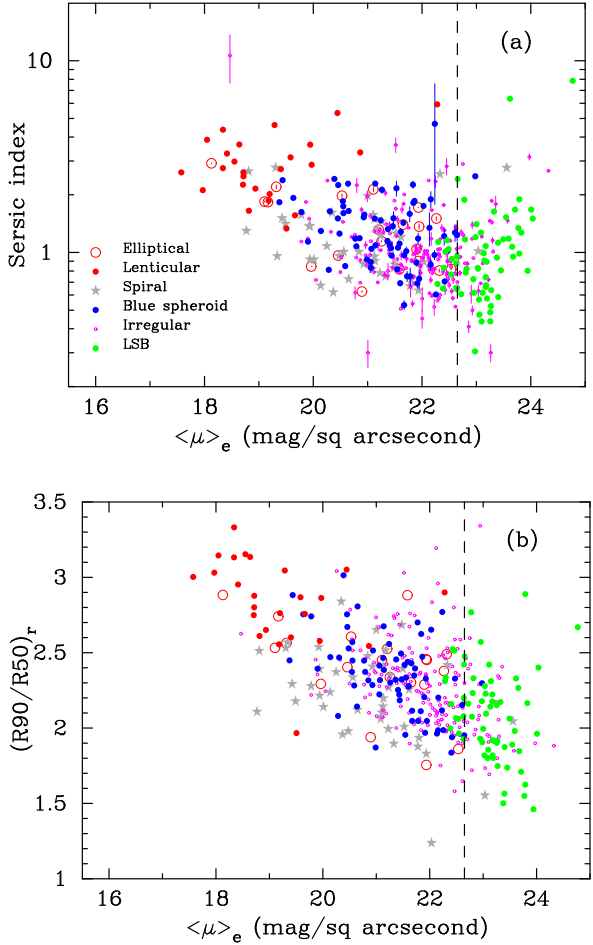

\centering{
 {\rotatebox{270}{\epsfig{file=figure5a.ps,width=6cm}}}}
\vspace{0.5cm}
\centering{
{\rotatebox{270}{\epsfig{file=figure5b.ps,width=6cm}}}}
\caption{(a) The S\'ersic index $n$ is shown as a function of the effective surface brightness \mue~for morphologically distinct galaxies
 in our \g II sample. Different morphological classes are represented by different colours and symbols as per the legend, and measurement
 uncertainties are shown for all classes except the LSBs to maintain clarity. The vertical dashed line represents the surface brightness
 completeness limit for our sample in the $Z$-band. (b) The same, but
 shown as a function of the concentration index $(R_{90}/R_{50})_r$. These figures show that albeit with a huge scatter, the effective surface
 brightness is correlated with the light concentration of galaxies, and the correlation between $n$ and \mue~seen in the upper panel
 is not a result of parameter coupling in the S\'ersic model. }
\label{sb-ser}
\end{figure}

 In order to test if the visual morphology of a galaxy is associated with its light profile, in Figure~\ref{sb-ser}
 we show the distribution of different morphological classes in surface brightness and S\'ersic index.
 Although there is almost an order of magnitude of scatter in $n$ at any \mue, a trend is clearly visible, such that
 higher surface brightness galaxies have higher stellar concentration towards the centre. In the bottom panel of
 Figure~\ref{sb-ser}, we show the same replacing $n$ with the concentration index $(R_{90}/R_{50})_r$, where
 $R_{x}$ is the Petrosian radius containing $x\%$ of the galaxy's light in the $r$-band. Together, these figures show that
  the correlation between $n$ and \mue~does not result from parameter coupling in the S\'ersic model.
  
 \begin{figure}
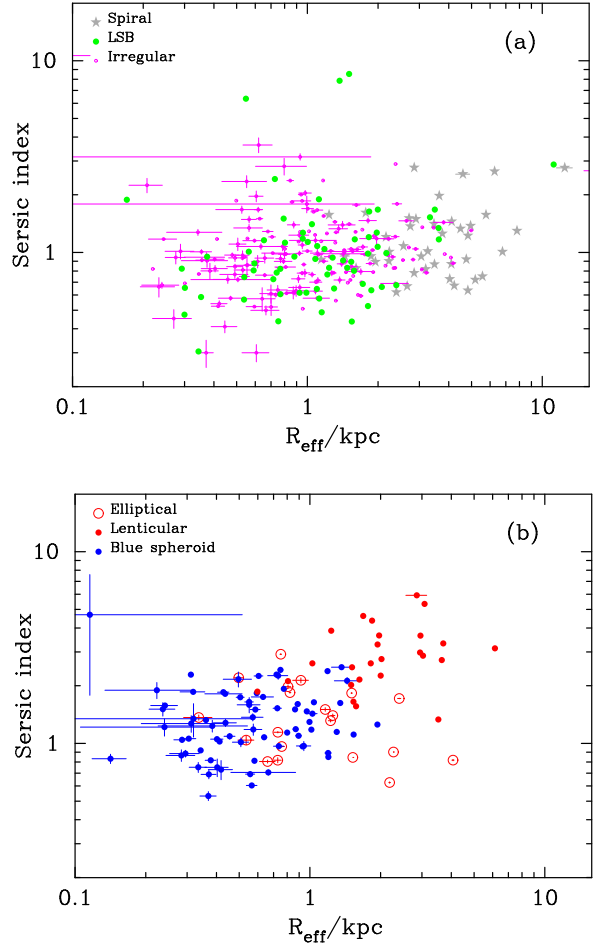

\centering{
{\rotatebox{270}{\epsfig{file=figure6a.ps,width=6cm}}}}
\vspace{0.5cm}
\centering{
{\rotatebox{270}{\epsfig{file=figure6b.ps,width=6cm}}}}
\caption{The S\'ersic index $n$ as a function of the effective radius \re~for (a) disk-dominated, 
and (b) bulge-dominated galaxies, respectively.
 Unlike the $n$-log \re~correlation presented in the literature (see text), and seen here for the bulge-dominated galaxies, we do
 not find any apparent trend between log \re~and $n$ for non-spheroidal galaxies.  }
\label{r-n}
\end{figure}

 Several authors have suggested a correlation between the physical radius and $n$ for luminous elliptical and lenticular galaxies,
 as well as dEs in clusters \citep[\eg][]{caon93,caon95,young95}. Even though our sample does not include classical $n=4$ ellipticals,
 in Figure~\ref{r-n} we find a similar correlation for the bulge-dominated galaxies in this sample. 
 It is particularly interesting that the star-forming blue spheroids form a continuous distribution with the passively evolving lenticulars
 in the n-\re~plane such that the product moment correlation $r=0.414$, while including ellipticals slightly worsens the correlation
  with $r=0.365$ (Figure~\ref{r-n}(b)).  
 Figure~\ref{r-n}(a) shows that although spiral galaxies are somewhat disjoint from the rest of
 the ensemble, none of the disk-dominated categories (Spirals, Irregulars and LSBs) show any apparent trend in this parameter space
 ($r=0.134$).
 Some of the intrinsic scatter within each morphological class may be due to the measurement uncertainties, or the incapability of the
 single component S\'ersic model to fully represent the stellar distribution in low-mass, faint galaxies represented by the irregulars,
 blue spheroids and LSBs. However, since the fraction of galaxies in each class where the light profile is not estimated properly is 
 small (Table~\ref{fit-stats}), we claim that the lack of a correlation between \re~and \mue~for non-spheroidal galaxies is real.
 The limitations of our data and methodology are discussed further in \S\ref{limit}.
 
\begin{figure}
\centering{
{\rotatebox{270}{\epsfig{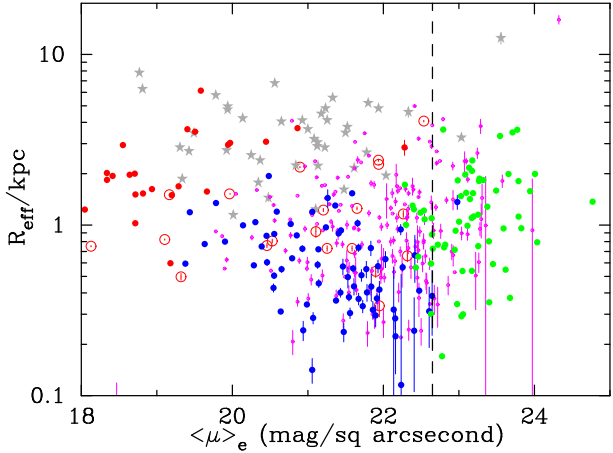}}}}
\caption{Effective radius, \re~as a function of the effective surface brightness within \re, \mue~for all 
 galaxies. The symbols and colours are same as in Figure~\ref{sb-ser}, and the vertical dashed line is the surface brightness
 limit for our sample. 
 The different locus of various morphological types shows the unconscious bias affecting visual classification.  }
\label{sb-r}
\end{figure}

 \begin{figure}
\centering{
{\rotatebox{270}{\epsfig{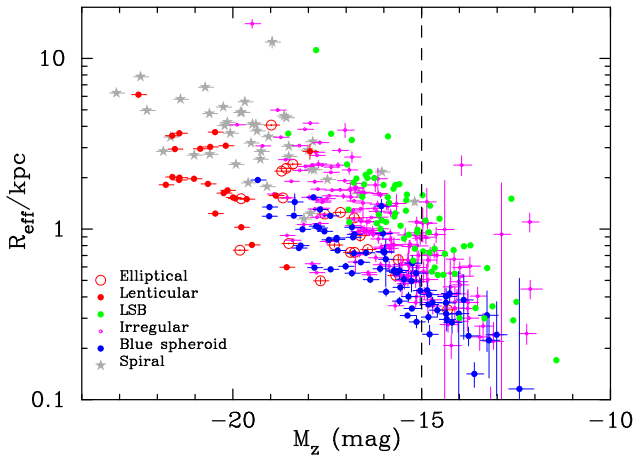}}}}
\caption{Effective radius, \re~as a function of the $Z$-band magnitude for galaxies in our sample. 
 The symbols and colours are same as in Figure~\ref{sb-ser}. The vertical dashed lines represents the limiting magnitude
 for our sample. This figure shows that albeit with some scatter, all galaxies show a monotonic relation between \re~and luminosity,
 such that the more luminous galaxies tend to be larger. At lower luminosities, LSBs tend to be larger than the blue spheroids and  the irregulars, while at the bright end, spirals have larger \re~relative to the lenticulars. }
\label{m-r}
\end{figure}  

 Figure~\ref{sb-r} shows the distribution of morphologically distinct galaxies in the plane spanned by effective surface
 brightness and the effective radius.
 As in Figure~\ref{sb-ser}, the different classes split into distinct populations, but with considerable overlap.  
 We quantify this observation using the Kolmogrov-Smirnov (K-S) statistical probability for various parameters for all pairs of 
 visually distinct galaxies in Appendix~\ref{ks:table}. Table~A5 for \re~shows that ellipticals are likely to have a statistically similar range
 in \re~to all types of galaxies except spirals and lenticulars. The LSBs and irregulars are also likely to have similar
 \re~distributions. Table~A8 on the other hand shows that the \mue~distributions are statistically different for all types of galaxies 
 except ellipticals, which show statistically similar distribution to spirals as well as blue spheroids. 
 Without the visual classification however, the sample fails to segregate into distinct population. 
 Such a lack of correlation between \mue~and \re~has previously been observed among luminous galaxies
  \citep[e.g.][]{boyce95}.
 
  Figure~\ref{m-r} shows the distribution of the galaxies in the $Z$-band magnitude and \re~plane.
 This figure adds strength to the above argument, by showing that the morphologically 
 distinct galaxies have a distinct locus in the 3D space mapped by absolute magnitude, surface brightness and effective
 radius, but {\it they all form a continuous distribution in this parameter space representing mass and angular momentum}. 
 
 \begin{figure}
\vspace{0.5cm}
\centering{
{\rotatebox{270}{\epsfig{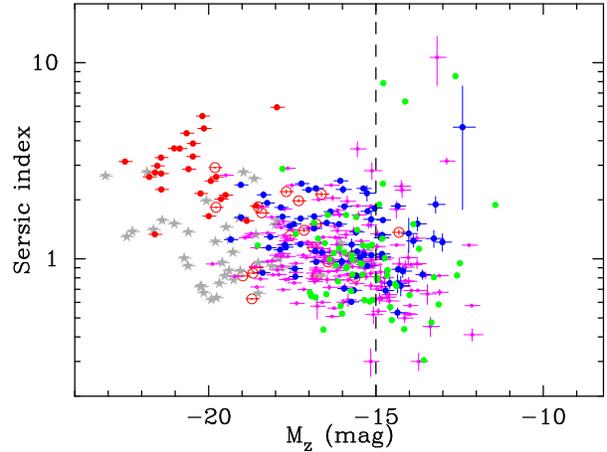}}}}
 \caption{S\'ersic index of galaxies as a function of $M_Z$. The vertical line marks the magnitude limit for our sample, and
 symbols and colours are same as in Figure~\ref{sb-ser}. While giant galaxies seem to segregate by morphological type,
 the low-luminosity galaxies form a unimodal population. It is noteworthy that the unimodality of the low-luminosity galaxies is not entirely a
 consequence of the measurement uncertainties in the estimated parameters, shown for all galaxies except the LSBs (see
  Table~\ref{means}).   }
\label{m-ser}
\end{figure}

 A similar trend continues in the plane spanned by absolute magnitude ($M_Z$) and $n$. As shown in Figure~\ref{m-ser},
 giant galaxies split into two branches, the passively evolving lenticulars at the high-$n$ end ($n>2$), and spirals at
 low-$n$ ($n\lesssim1$). The dwarfs on the other hand do not show any trend in this space with
 LSBs, irregulars and blue spheroids forming a single `clump' in the dwarf regime ($M_Z> -18$). The product moment correlation
 for the entire sample in this plane is $r=-0.210$, which reduces to a negligible $-0.065$ when only dwarfs are considered.
 The K-S statistic suggests that among dwarfs, 
 blue spheroids and ellipticals, and irregulars and LSBs show statistically similar distributions in $n$ (Appendix~\ref{ks:table}).
 The $n$ distribution of spiral galaxies is also statistically similar to irregulars and LSBs. 
 
 Together, Figures~\ref{sb-ser} to \ref{m-ser} show that while S\'ersic index is a good quantitative representation of the morphology
 of luminous galaxies, it is not an effective measure for dwarfs, even at the low redshift ($z\leq 0.02$) considered here.

\begin{figure}
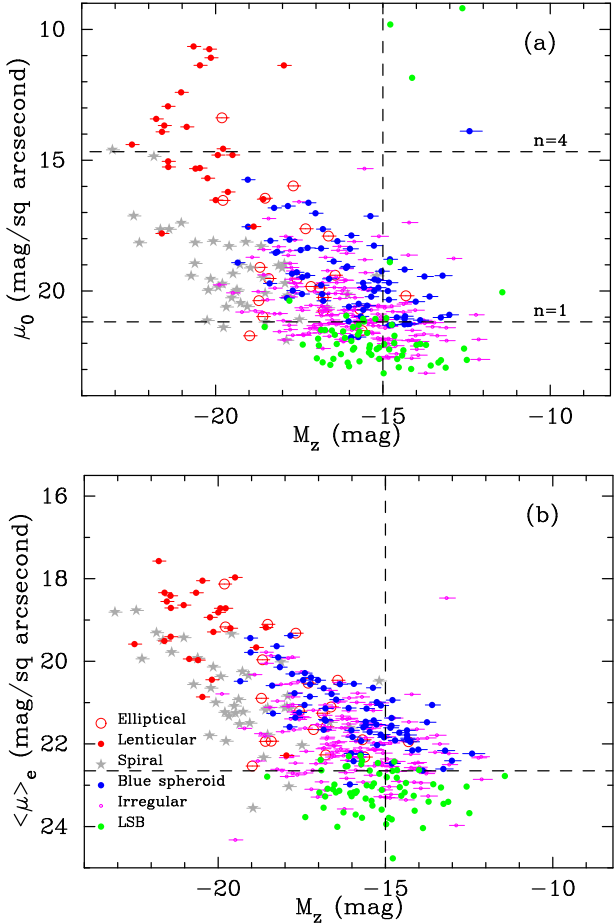

\centering{
{\rotatebox{270}{\epsfig{file=figure10a.ps,width=6cm}}}}
\vspace{0.25cm}
\centering{
{\rotatebox{270}{\epsfig{file=figure10b.ps,width=6cm}}}}
\caption{(a) Central surface brightness and (b) effective surface brightness, respectively as a function of
 $M_Z$ for our sample. The symbols and colours are as in the legend. The vertical dashed line marks the
 limiting magnitude for our \g II sample. For the central surface brightness {\it (top)}, the horizontal dashed lines represent the expected values
 corresponding to the S\'ersic index, n$=1$ and $4$ using the formulae derived by \citet{graham05}. In the {\it bottom} panel
 the horizontal line represents the surface brightness completeness limit for our sample. }
\label{m-sb}
\end{figure}

Figure~\ref{m-sb} shows the bivariate brightness distribution (BBD) for our sample in the $M_Z$-\mue($\mu_0$) planes.
It is remarkable how different morphological classes segregate into distinct spaces in the plane
mapped by the (central/effective) surface brightness and absolute magnitude, but link to form a continuum across the entire range.
This is reflected in the K-S statistical probabilities for different types of galaxies showing the distribution of \mue, $\mu_0$ and
 magnitude to be drawn from the same parent population. Appendix~\ref{ks:table} shows that the distributions of \mue~and $\mu_0$
 are different for all combinations of visually distinct galaxies except the ellipticals and blue spheroids, and, ellipticals and spirals.
 It is interesting that this trend is seen in the more commonly used $\mu_0$ and not just in \mue,
 because \mue~can be measured more robustly relative to $\mu_0$ whose measure is strongly affected by the quality of seeing
  \citep{graham03,boselli08}.
 
 The giant galaxies in Figure~\ref{m-sb} split into two sequences, one of star-forming spirals and the other formed by lenticulars.
 The lenticulars in particular have statistically distinctive distribution of \mue~relative to all other types of galaxies (Appendix~\ref{ks:table}). 
  At a given magnitude, spirals on an average have \mue~fainter by around
 two magnitudes per square arcsecond than the lenticulars, and $\mu_0$ fainter by around four magnitudes per square arcsecond,
 respectively. 
 
    \begin{table*}
\caption{Mean (standard deviation) and the standard error on mean of the distribution of structural parameters for different morphological classes.}
\begin{tabular}{|c|c|c|c|c|c|c|}
\hline 
 Class $\rightarrow$           & Elliptical & Lenticular & Spiral & BSph & Irr  & LSB \\
 Parameter $\downarrow$ & & & & & & \\ \hline \hline
$\mu_0$ (mag/sq arcsecond) & $17.95(5.45)\pm1.19$ & $14.20(2.02)\pm0.40$ & $19.25(1.51)\pm0.22$ & $19.10(3.20)\pm0.37$ & $20.18(3.95)\pm0.28$ & $21.15(3.45)\pm0.41$ \\ 
\mue (mag/ sq arcsecond) & $22.06(1.64)\pm0.36$  & $20.37(1.07)\pm0.21$  & $21.59(1.02)\pm0.15$  & $22.19(0.96)\pm0.11$  & $22.63(1.01)\pm0.07$  & $23.79(1.31)\pm0.16$  \\
\re (kpc) & $5.21(4.11)\pm0.90$  & $9.43(6.56)\pm1.29$  & $19.91(12.24)\pm1.78$  & $3.36(2.68)\pm0.31$  & $6.03(5.05)\pm0.36$  & $5.74(4.30)\pm0.52$  \\
$n$ & $2.05(2.95)\pm0.64$  & $3.01(1.13)\pm0.22$  & $1.24(0.56)\pm0.08$  & $1.59(1.70)\pm0.20$  & $1.29(1.68)\pm0.12$  & $1.37(1.57)\pm0.19$  \\
 \hline
\end{tabular}
\label{means}
\end{table*}

 To summarise, in this section we have shown that although morphologically distinct galaxies have different centroids and statistically
 different distributions in many of the 2D parameter spaces mapped by luminosity and the structural parameters, the different classes
 especially among the dwarf galaxies overlap significantly. 
 Table~\ref{means} shows the mean, standard deviation and the standard error on mean for \mue, $\mu_0$, \re~and $n$,
 respectively for different morphological types. Table~\ref{means} shows that despite being empirically distinctive, the
 distributions of structural parameters of visually distinct galaxies are very broad and overlap extensively, even though the  measurement uncertainties are negligible relative to the scatter within each class. 
 The broad width of the distributions, specifically of \re~and $\mu_0$ (i) may be an indication of different evolution histories of
 morphologically similar galaxies, or, (ii) a result of an artificial categorisation of a continuous galaxy morphology distribution into
 discontinuous classes.
 The same however may also indicate (i) the inefficient modelling of the light profile 
 for some structural parameters such as $\mu_0$ (Table~\ref{fit-stats}; also see \S\ref{limit}), and, (ii) uncertainty in our visual
 classification of the low-luminosity galaxies. Better imaging data and multi-component modelling of light profile are required to
 confirm this.
 
\section{Star formation and dust in morphologically distinct galaxies}
\label{sf}

With the aim of getting further insight into the differences between morphologically distinct galaxies, we fitted the
 multi-band photometry available for most of our sample, with {\sc magphys} (\S\ref{mag}). 
 Here we describe how morphologically distinct galaxies behave in the parameter space 
formed by the star formation and dust properties of galaxies.  
\begin{figure*}
\centering{
{\rotatebox{270}{\epsfig{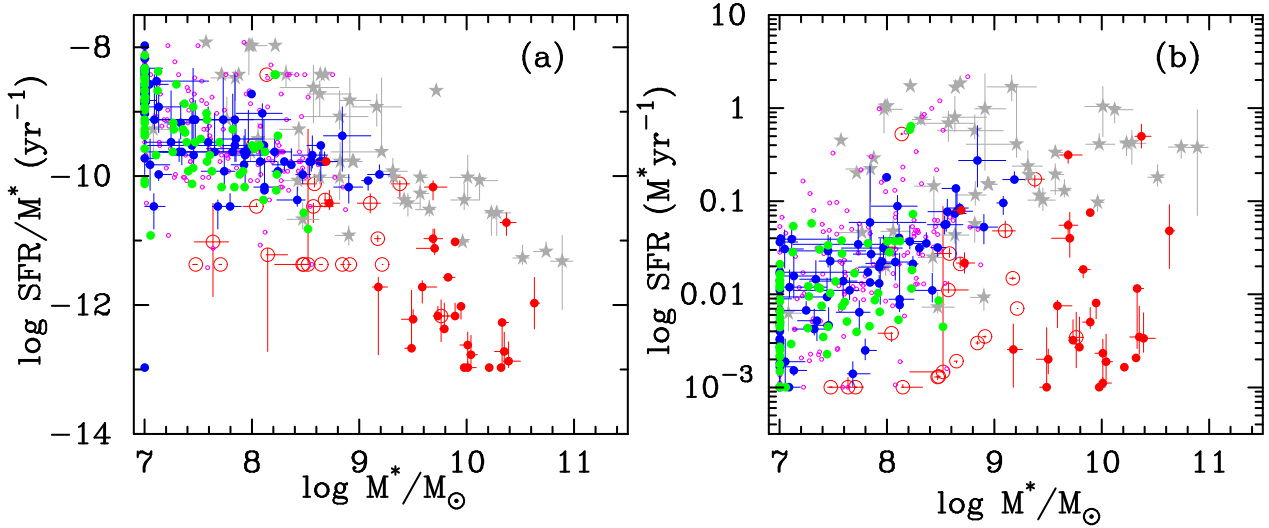}}}}
\caption{(a) sSFR and (b) SFR of different types of galaxies as a function of log $M^*$. The symbols and colours
are the same as in Figure~\ref{sb-ser}.
 All parameters shown here are the medians of the PDFs obtained for the corresponding parameter 
 from the spectral energy distribution fitting using {\sc magphys} (see text). The lower and upper uncertainties
 represent the 16th and the 84th percentiles of the PDFs, respectively.  }
\label{sf-lgm}
\end{figure*} 
 
\begin{table*}
\caption{Mean (standard deviation) and the standard error on mean of the distribution of star formation and dust properties for different morphological classes.}
\begin{tabular}{|c|c|c|c|c|c|c|}
\hline 
 Class $\rightarrow$           & Elliptical & Lenticular & Spiral & BSph & Irr  & LSB \\
 Parameter $\downarrow$ & & & & & & \\ \hline \hline
log SFR ($M_\odot yr^{-1}$) & $-2.20(0.77)\pm0.17$  & $-2.03(0.75)\pm0.15$  & $-0.65(0.64)\pm0.09$  & $-1.83(0.57)\pm0.07$  & $-1.74(0.69)\pm0.05$  & $-2.05(0.56)\pm0.07$  \\
log SFR/$M^*$ (yr$^{-1}$) & $-10.84(0.81)\pm0.18$  & $-11.89(0.93)\pm0.18$  & $-9.60(1.00)\pm0.14$  & $-9.42(0.73)\pm0.08$  & $-9.28(0.71)\pm0.05$  & $-9.30(0.66)\pm0.08$  \\
log $M^* (M_\odot$) & $8.55(0.58)\pm0.13$  & $9.86(0.47)\pm0.09$  & $8.95(0.92)\pm0.13$  & $7.68(0.64)\pm0.07$  & $7.63(0.54)\pm0.04$  & $7.41(0.45)\pm0.05$  \\
log $M_{dust} (M_\odot$) & $5.02(0.64)\pm0.14$  & $6.06(0.65)\pm0.13$  & $6.41(0.75)\pm0.01$  & $4.74(0.66)\pm0.08$  & $4.85(0.65)\pm0.05$  & $4.64(0.64)\pm0.08$  \\
 \hline
\end{tabular}
\label{sf-means}
\end{table*}
 
 Figure~\ref{sf-lgm} shows the well-known trend in the star formation rate (SFR) with $M*$, viz., SFR increases, while specific SFR,
 SFR/$M^*$ (sSFR henceforth) decreases with increasing stellar mass of galaxies, respectively. In this sample spanning almost four
 orders of magnitude in $M^*$, once again the star-forming and passive giant galaxies occupy very different ranges in both, $M^*$
 and sSFR. Since our sample does not include clusters (\S\ref{data}),
 it is unsurprising that the visually identified elliptical galaxies are mostly low-mass ellipticals ((d)ellipticals, henceforth) and 
 not the classic giant ellipticals dominating the cores of rich clusters. 
    
 As an ensemble all dwarf galaxies ($M^*/M_\odot \lesssim 10^9$) show steadily increasing SFR with $M^*$. The K-S statistic
 suggests that the distribution of $M^*$ for all except the irregulars and blue spheroids, and SFR for all except the lenticular and (d)ellipticals,
 and lenticulars and LSBs, respectively are statistically different (Appendix~\ref{ks:table}). The sSFR distribution on the other hand
 is similar for the irregulars, blue spheroids and the LSBs, thereby showing a unimodality among the star-forming dwarf galaxies. 
 Some of this overlap in the star formation properties of dwarf galaxies may however be a result of the scatter within each
 morphological class as suggested by the mean and standard deviations for the distributions of SFR, sSFR, $M^*$ and $M_{dust}$
 for visually distinct galaxies (Table~\ref{sf-means}). 
 
\begin{figure}
\centering{
{\rotatebox{270}{\epsfig{file=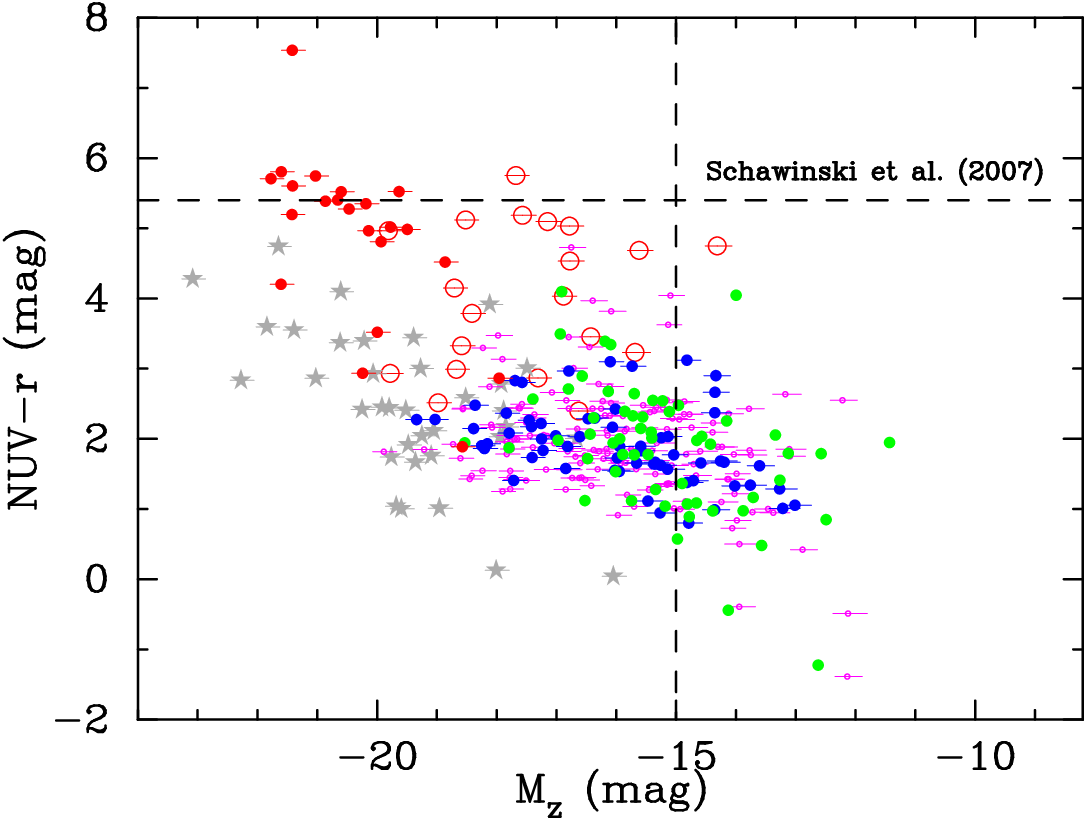,width=6cm}}}}
\caption{\nr~colour as a function of the $z$-band absolute magnitude for all galaxies in our sample. The symbols and colours
are the same as in Figure~\ref{sb-ser}. The dashed horizontal line at \nr=$5.4$ is the lower limit
for the passively evolving early-type galaxies in the SDSS \citep{schawinski07}, while the vertical dashed line marks the magnitude
completeness limit of our sample. This figure shows that the passively-evolving lenticulars are on average around $2.5$ magnitude
 redder than the spiral galaxies of similar luminosity. In the low-luminosity regime however, all dwarf galaxies except the
 ellipticals form a unimodal population. The ellipticals on average are redder than all other dwarfs of similar luminosity.  }
\label{cc}
\end{figure}

 If a galaxy has spent most of its life away from the influence of cluster and large groups,
 it may be assumed that the phase of its passive evolution begins once it runs out of all the cold gas which would have fuelled star formation.
  Under these assumptions, Figure~\ref{sf-lgm} qualitatively indicates that large spiral galaxies such
 as those probed here, turn into passively evolving lenticulars, while the low-mass galaxies may turn into small red (d)ellipticals.
 This observation is based on the distribution of galaxies in both panels of Figure~\ref{sf-lgm}, where
 the (d)ellipticals appear in the same mass range as star-forming dwarf galaxies, but with extremely low star formation. 
 
 Figure~\ref{cc} shows the distribution of our galaxies with UV data on the $(NUV-r)$ versus $M_z$ colour magnitude diagram.
  The UV-optical colour of (d)elliptical galaxies shows that at fixed luminosity, they are on average
 redder than their counterparts, while all the other dwarf galaxies form a unimodal population. In the high luminosity regime 
 on the other hand, the lenticulars and spiral galaxies clearly segregate, such that the lenticulars are on average around 2.5
 magnitude redder than the spiral galaxies. This figure further strengthens the above hypothesis that a star-forming dwarf galaxy (LSB, irregular or blue spheroid) may evolve into red (d)elliptical, while a spiral
  galaxy will turn into a lenticular.
 
\begin{figure*}
\centering{
{\rotatebox{270}{\epsfig{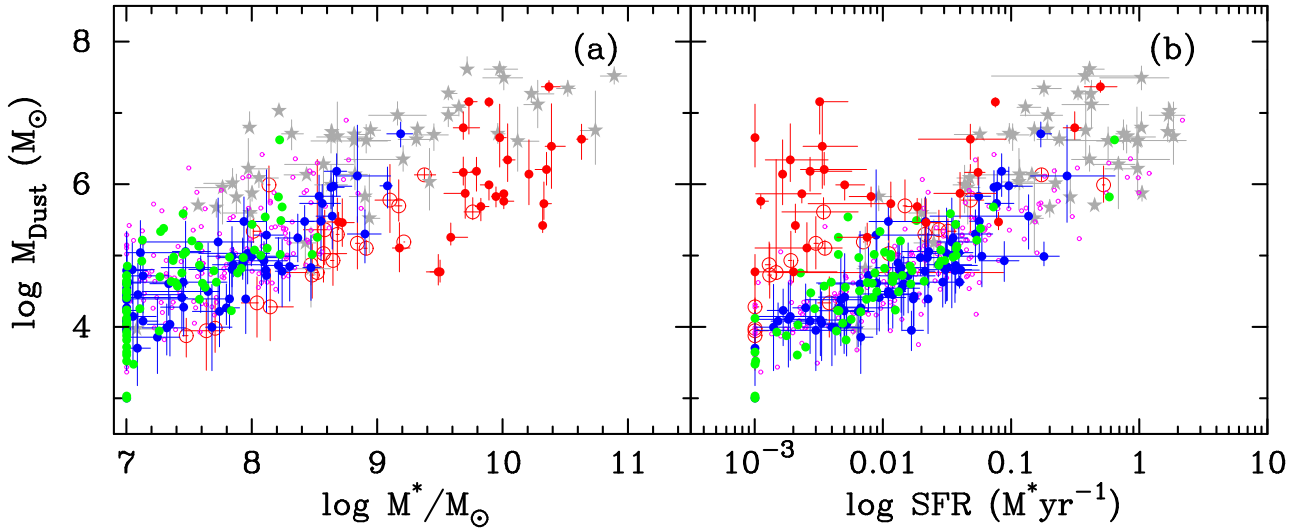}}}}
\caption{(a) Dust mass in different types of galaxies as a function of $M^*$ and (b) the total SFR,
respectively. All quantities plotted are determined from spectral energy distribution fitting using {\sc magphys} (see text).
The symbols and colours are the same as in Figure~\ref{sb-ser}. The plotted values are the median, while uncertainties are the 16th and the
 84th percentile values obtained from the probability distribution function for all the parameters for each galaxy, respectively. }
\label{dust}
\end{figure*}

In Figure~\ref{dust} we analyse the dust mass of galaxies in our sample as a function of $M^*$ and the total SFR, respectively.
Although with considerable scatter, the sample shows a trend of increasing dust mass $M_{dust}$, with $M^*$,
and SFR. As expected, spiral galaxies have the highest SFRs and dust masses at fixed $M^*$ (also see Table~\ref{sf-means}).
 The lenticulars on the other
hand, have at least an order of magnitude lower $M_{dust}$ at fixed $M^*$, and several orders of magnitude lower SFRs on average.
 K-S test probabilities (Appendix~\ref{ks:table}) suggest that the distribution of $M_{dust}$ for (d)ellipticals is statistically similar
 to irregulars and blue spheroids, while that of the latter two is similar to LSBs as well. 
 
The star-forming dwarf galaxies; irregulars, blue spheroids and LSBs, show a monotonic increase in $M_{dust}$ with $M^*$ and
SFR, but no apparent trends segregating different classes. At fixed $M^*$, (d)ellipticals have lowest $M_{dust}$. This trend however
reverses with SFR, such that at fixed SFR (d)ellipticals have the highest dust masses relative to other dwarf classes. In the
 SFR-$M_{dust}$ space, (d)ellipticals mark the lower limit of passively evolving galaxies in $M_{dust}$.
 
 If we assume that dust is not easily destroyed in galaxies, and that galaxies can not produce dust once their star formation turns off,
 the trend seen in Figure~\ref{dust}(a) suggests that the dwarf galaxies will turn into (d)ellipticals as they age.
 On the other hand if $M_{dust}$ in the giant spiral galaxies is fixed when star formation turns off, they will fade into the lenticular
 class identified here. The fact that $M_{dust}$ of lenticulars is on average a magnitude lower than the spirals does not contradict
 this hypothesis because star-forming galaxies tend to loose dust via outflows such as supernovae explosions during their
 ``active" phase. Therefore the total dust mass of a galaxy towards the end of its star-forming phase is likely to be less than at the
 peak of its star formation activity.   
 These observations thus imply that star-forming dwarf galaxies may have evolved into (d)ellipticals, while lenticulars succeed
  star-forming spirals. 

To summarise, the star formation and dust properties derived from {\sc magphys} efficiently separate giant galaxies into spirals
 and passively evolving lenticulars, but fail to distinguish the morphologically distinct star-forming dwarf classes. 
 This is likely a consequence of intrinsic scatter within each class and not the measurement uncertainties as indicated by
 Table~\ref{sf-means}.
 The (d)ellipticals on the other hand, span a distinguishable range of SFR and sSFR relative to other galaxies of similar luminosity.

\section{Discussion}
\label{discussion}
\subsection{Stellar mass distribution for different morphological types and a comparison with the cluster environment}

\begin{figure}
\centering{
{\rotatebox{270}{\epsfig{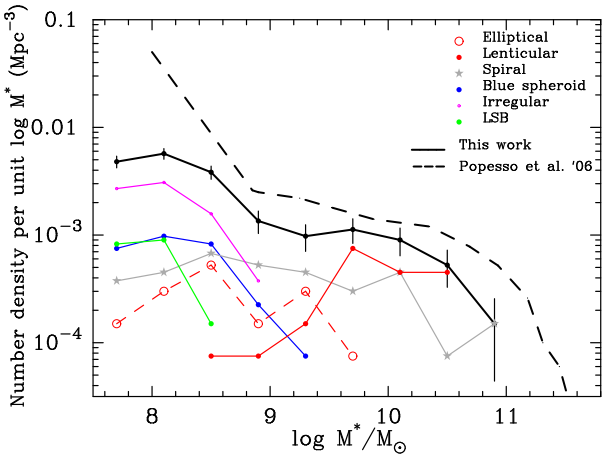}}}}
\caption{The stellar mass function of galaxies derived for individual morphological classes, the composite
 for the entire sample, and the composite of cluster galaxies based on the luminosity function derived by \citet{popesso06} and converted to
 $M^*$ by \citet{baldry08}. The mass function for cluster galaxies was scaled down by a factor of five to aid in comparison. 
 Poissonian uncertainties are shown for the composite mass function for \g II sample. While an upturn
 at $log M^*<9$ is apparent in both samples, the low-mass end in clusters rises more sharply compared to the
 environments probed by our GAMA sample. It is also evident that the low-mass branch in the \g sample is mainly populated by the LSBs, Irregulars and blue spheroids, while
 in clusters dEs are more prominent at this end. }
\label{mass}
\end{figure}

 In Figure~\ref{mass} we show the stellar mass function for all the galaxies in our sample, function corresponding to different
 morphological classes and a representative function for cluster galaxies. 
 The cluster mass function was presented by \citet[][see their figure~7]{baldry08}
 for clusters in the RASS-SDSS galaxy cluster survey \citep{popesso06}. Using the relation given by \citet{taylor11}
  for our sample complete to $M_Z=-15$ at $z=0.02$, the stellar mass completeness ranges from $7.15\leq$log$M^*\leq7.99$,
 corresponding to $0\leq(g-i)\leq1.2$. The upturn at log$M^*<9$ seen in clusters
 \citep[also see][and references therein]{jenkins07,yamanoi12} is apparent in our
  \g sample as well, consistent
 with other comparisons of galaxy stellar mass function in different environments \citep{baldry08}. 
 The low-mass upturn in clusters is evidently steeper than for our \g  mass function, suggesting that the slope of the mass
 function at the low-mass end is environment-dependant. Note, however, that the cluster distribution is derived by statistical subtraction of background galaxy counts. 
 
 There is also a marked difference in the galaxy populations which contribute to the low-mass branch of the mass function. It is noteworthy
 that in the environments probed here, LSBs, blue spheroids and irregulars, all of which are forming stars (Figure~\ref{sf-lgm}),
 dominate at log$M^*<9$, while in clusters this branch is mostly populated by passively evolving dEs. Deep,
 spectroscopic data homogeneously sampling different environments are thus required to confirm the differences seen in the faint-end
 slope of the mass function for different environments. 
 
\subsection{The bivariate brightness distribution: A comparison with literature}

 Generically speaking, galaxies should be describable using two independent parameters \citep{brosche73,kodaira83}.
 While several combinations of such parameters have been explored in literature
 \citep[\eg][]{caldwell83,kodaira83,binggeli84,kodaira89,young95,prugniel97},
 the bivariate brightness distribution (BBD) appears to be the most enlightening in the context of structural evolution of galaxies
 during their life cycle. The BBD showing the natural morphological segregation of Virgo cluster galaxies in the space mapped by
 central surface brightness and absolute magnitude has been shown by several authors \citep{caldwell83,binggeli84,kormendy85,binggeli94}.
 From these seminal works, in Binggeli's words, ``The $M-\mu$ diagram for stellar systems might well become the equivalent of the HR
 diagram for stars".  

 In this work, the BBD in its two forms comparable to Binggeli's figures is shown in Figure~\ref{m-sb}. 
 The first conclusion that can be drawn from these figures
 is that the effective surface brightness within the effective radius is directly correlated with luminosity. This is an intuitive, yet non-obvious
 inference. In other words, low-luminosity, high surface brightness galaxies
 \citep[e.g. M60-UCD1;][]{strader13}, and luminous, low surface brightness galaxies like Malin I \citep{bothun87} are very rare.    
   
 In the environments probed here, the giant galaxies split into two categories such that
 the high surface brightness branch comprises red and passively evolving lenticular galaxies (Figures~\ref{sf-lgm} and \ref{cc}).
 The low surface brightness branch on the other hand is made up of star-forming spiral galaxies. 
 The dwarfs  do not show a trend with luminosity; rather each class of dwarf galaxies spans the same range of luminosity
 but the effective surface brightness increases between classes, in increasing order: LSBs, the irregulars and the blue spheroids
 (also see Table~\ref{means}).
 This trend is visible with $\mu_0$ and \mue, although the segregation and mild trend with luminosity is more obviously seen with
 the latter.
 
 Most of the literature following the papers published in the early 1980s either concentrated on resolving the dichotomy between giant
 and dwarf elliptical galaxies \citep[e.g.][and references therein]{graham03,kormendy12,graham13}, or the evolutionary sequence from
 bulges of spiral galaxies to dwarf ellipticals \citep[\eg][and references therein]{boselli08,toloba11}.
 But since dwarf ellipticals have predominantly been discovered in nearby rich clusters such as Virgo and Coma,
 all such papers are somewhat related to the impact of environmental processes in clusters on the structural properties
 of galaxies. Our work in this paper, on the other hand, gives a complimentary view of galaxy properties, particularly dwarfs, 
 in environments outside rich clusters.
   
  \subsection{On the issue of statistically distinct populations of galaxies in the local Universe}
 
 \begin{figure}
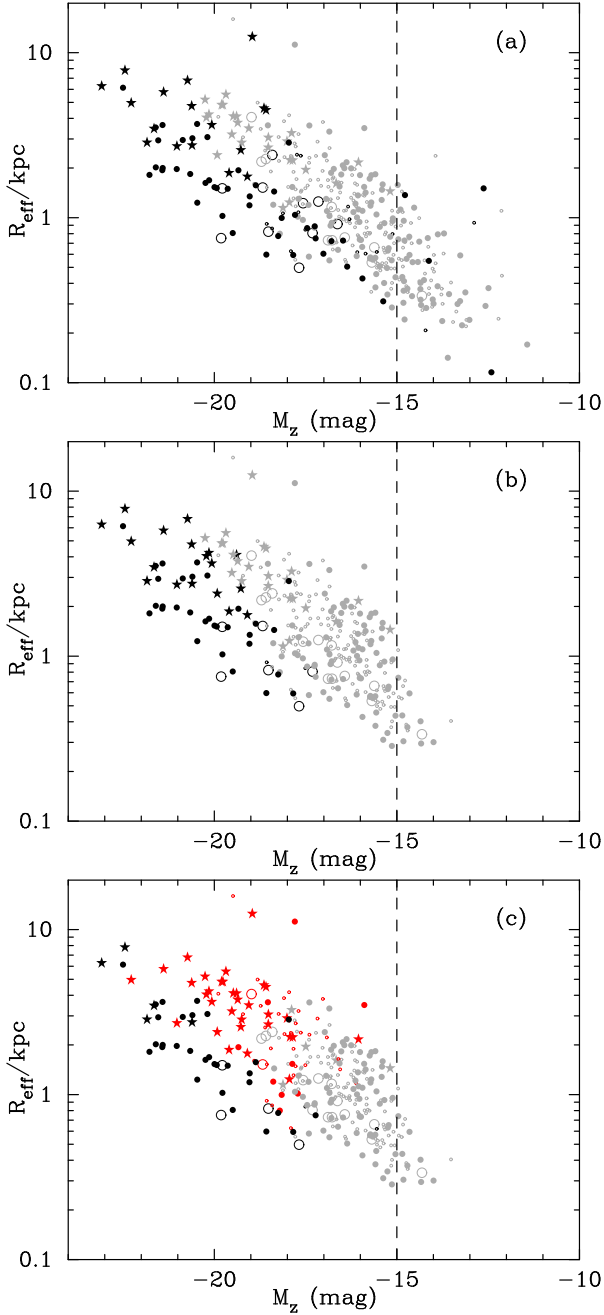

\centering{
{\rotatebox{270}{\epsfig{file=figure15-3a.ps,width=5.8cm}}}}
\centering{
{\rotatebox{270}{\epsfig{file=figure15-3b.ps,width=5.8cm}}}}
\centering{
{\rotatebox{270}{\epsfig{file=figure15-3c.ps,width=5.8cm}}}}
 \caption{Automated classification of galaxies in our sample. This is the same as Figure~\ref{m-r}, 
 but colour-coded for ``clusters'' identified by the k-means algorithm using different combinations
 of parameters: (a) Structural (b) Structural+SF 1, and (c) Structural+SF 3, respectively as discussed in the text (also see
 Table~\ref{clusterlist}). The symbol types are same as in the above figures.  
 Two clusters are preferred statistically for this sample for all but one combination of parameters. There is however no
 statistical evidence for any partition among the low-luminosity galaxies.  }  
 \label{kmeans}
 \end{figure}

One of our primary aims was to objectively test how many distinct classes are required to classify the nearby galaxy population, as opposed to the number of classes we inferred from our visual classifications. As shown by the 2-dimensional parameter plots in \S\ref{photo} and \S\ref{sf}, galaxies with different visual classifications have overlapping distributions at fainter luminosities, but appear to split into star-forming and passively-evolving classes at brighter luminosities. It is still possible that the dwarf galaxies separate into different clusters in the full multi-dimensional parameter space that may have been missed by the two-dimensional projections. We therefore applied a quantitative statistical analysis to test our data for evidence of distinct groups in the multi-dimensional parameter space.

\subsubsection{Application of the Clustering Algorithm}
\label{s:nclus}

We used the ``k-means'' algorithm \citep{macqueen67} to decompose the data into 2-20 clusters. For a given number of clusters, this finds the cluster positions that minimise the sum of the squares of the distances from each data point to its cluster centre. We determined the best number of clusters to choose by using the \citet{calinski74} variance ratio criterion \citep[as implemented in the NbClust package][]{charrad13}. This involves choosing the partition that gives the highest ratio of the variance of the distances between objects in different clusters to the variance of distances between objects within clusters. \citet{calinski74} plot the variance ratio as a function of the number of clusters and use the first local maximum to define the best number of clusters. We modified this criterion by requiring that the selected peak was significantly (3 standard deviations) above the neighbouring points\footnote{The NbClust package \citep{charrad13} does not take into account uncertainties in the parameters and does not calculate uncertainties in the variance ratios, so we used a Monte Carlo approach to estimate uncertainties. We recalculated each variance ratio 30 times using different initial random seeds for the k-means algorithm and used the resulting distributions to estimate the uncertainties: see Appendix~\ref{nclusters}.}. On rare occasions a double peak was formed by two variance ratios with the same value (to within 1 standard deviation) that was still significantly higher than the neighbouring points. In this case the first of the two (always $n=2$ clusters) was chosen.

Before starting the clustering analysis we removed any objects with missing data and then scaled the remaining galaxies to have a mean of zero and standard deviation of unity in each parameter. All the parameters we analysed were logarithmic measurements (or magnitudes): the algorithm failed to find a preferred number of clusters when applied to linear units, apparently because the distribution in each parameter consisted of a group plus one or two extreme outliers.           
           
\subsubsection{Clustering with Structural Parameters}

We first applied the analysis to the four structural parameters from {\sc sigma} ($\mu_0$, \mue, $\log n$ and \re) and the optical ($g-i$) colour giving a sample of 405 galaxies. We define this as our reference data set as it is closest to the information we used for our visual classifications. For these parameters, two clusters were preferred over any other partition of the data. This partition is summarised (as ``Structural'') in Table~\ref{clusterlist} and illustrated in Figure~\ref{kmeans}(a). The figure is same as the $M_Z$-\re~distribution shown in Figure~\ref{m-r} above, but colour-coded by the clusters identified by the k-means algorithm. The first cluster (91 galaxies) tends to higher luminosity and the second cluster (314) has lower luminosities. Where the two clusters overlap in luminosity, objects in the first cluster have smaller radii as shown in Figure~\ref{kmeans}(a). In terms of our morphological classification, the first cluster contains all the ``lenticulars'' as well all the more luminous spirals.
 It also contains some of the more luminous ``elliptical'' and ``blue spheroid'' galaxies.

\begin{table*}
\caption{Automated classification results with different choices of parameters}
\label{clusterlist}
\vspace*{3mm}
\begin{tabular}{llrrrrrr}
\hline
partition & parameters   & N & $n_1$& $n_2$& $n_3$ & $p_{Vis}$ & $p_{-18.5}$  \\
\hline
Structural & 12345-{-}{-}{-}  & 405&  91  &  314 &  -  &   88\%&   87\%         \\
Structural+SF 1 & 123456789        & 313&  57  &  256 &  -  &   97\%&   97\%         \\
Structural+SF 2 & 1234-6789        & 313&  59  &  254 &  -  &   97\%&   97\%         \\
Structural+SF 3 & 1234-67-9        & 313&  41  &  198 & 74  &   79\%&   83\%         \\
\hline
\end{tabular}
\vspace*{3mm}

Note: the parameters used in each choice are numbered as follows. 
1: $\mu_0$; 
2: \mue; 
3: $\log n$; 
4: $\log $\re; 
5: $g-i$; 
6: $\log sSFR$; 
7: $\log SFR$; 
8: $\log M^*$; 
9: $\log M_{dust}$. 
$N$ is the number of galaxies and $n_i$ are the number of galaxies assigned to cluster $i$ as colour-coded in Figures~\ref{kmeans}(a)-(c). $p_{Vis}$ is the percentage of visually-classified star-forming dwarf galaxies assigned to the second cluster and $p_{-18.5}$ are the percentages of galaxies with absolute magnitude $M_Z>-18.5$ in the second cluster. 
\end{table*}

We investigated the dependence of the partitions on the parameter choice by removing different parameters from the clustering calculation. Removing the colour had very little effect (2 clusters of 97 and 308 members), but when we removed the colour and any one of the structural parameters, the k-means algorithm could not find any preferred partition. We then retained the colour but removed each of the structural parameters in turn. Again, these gave very similar results, except when either of the surface brightness terms was removed. The size of the first cluster increased to 129 when the central surface brightness was removed ($\mu_0$) and it decreased to 78 when the effective surface brightness (\mue) was removed. These changes reflect the properties of the respective surface brightness measures shown in Figure~\ref{m-sb}. The central surface brightness very strongly separates the different luminous galaxy classes, so removing it causes the clustering algorithm to tend to put all the luminous galaxies together. The converse applies to the effective surface brightness.

We observed two common features of all the partitions based on the structural plus colour parameters. First, the lenticular galaxies were always all included in the first (high-luminosity) cluster, as were the luminous ($M_Z<-20.5$) spirals. Second, there was no separation of the low-luminosity galaxies into separate classes. Instead, the majority (88\%) of the objects we visually classified as star-forming dwarfs (BSph, Irr, LSB) were placed in the second cluster. The fraction of {\em all} low luminosity ($M_Z>-18.5$) galaxies in the second cluster is 87 per cent.


\subsubsection{Clustering with Structural and {\sc magphys} Parameters}

We then added the four {\sc magphys} parameters (log of sSFR, SFR, $M^*$ and $M_{dust}$), creating a nine-dimensional space (but with degeneracy between some of the parameters). This sample was limited to 313 galaxies due to incomplete {\sc magphys} data. This also gave a preference for two clusters but with fewer galaxies (57) in the high-luminosity group than with just the structural parameters. This is shown in Figure~\ref{kmeans}(b) and listed as ``Structural+SF 1'' in Table~\ref{clusterlist}. The main difference from the previous partition is that very few low luminosity galaxies now appear in the first cluster. The fraction of galaxies in the second cluster is now 97 per cent at luminosities fainter than $M_Z=-18.5$. The first cluster consists of just the visually-classified lenticulars and the brighter spirals. The second cluster contains 97 per cent of all the visually-classified star-forming dwarf galaxies.

As previously, we removed the ($g-i$) colour and found that this made no significant difference to the result (59 in the first cluster; ``Structural+SF 2'' in Table~\ref{clusterlist}). This is not surprising as the {\sc magphys} parameters are based on extensive colour information; indeed $g-i$ correlates strongly with specific star formation ($R=-0.72$). For the remaining tests we therefore excluded the $g-i$ colour. We then removed each of the four {\sc magphys} parameters in turn. Removing any one of specific SFR, SFR and dust mass still resulted in two clusters, but with varying numbers of spirals in the first cluster (cluster 1 sizes 93, 56, and 47 respectively). Removing the stellar mass, by contrast, resulted in a significant preference for a 3-cluster partition which we show in Figure~\ref{kmeans}(c) (``Structural+SF 3'' in Table~\ref{clusterlist}). The first cluster (high luminosity and low radius) is now smaller and corresponds very closely to the visual lenticular classification. The remaining two clusters contain all the other galaxies split into high- and low-luminosity groups (74 and 198). 

\subsubsection{Summary of clustering analysis}

Clustering algorithms like k-means provide an objective basis to determine how many distinct groups are defined by our measured galaxy parameters. Our analysis has shown that the results are somewhat sensitive to the choice of parameters, notably at higher luminosities where the numbers of galaxies in the first cluster can vary by up to 30 per cent. However at lower luminosities ($M_Z>-18.5$) there was much less variation: in the two-cluster partitions the number of  galaxies in the second cluster varied by only 5 per cent. 

In all but one of the partitions we have discussed, only 2 clusters were identified in the data. This is markedly different from our visual classifications which identified 6 different galaxy types. The most striking result is that the three visual classes of dwarf star-forming galaxies (blue spheroid, irregular and low surface brightness) are assigned to the same cluster. Using just the structural parameters 88 per cent of the star-forming dwarfs are in cluster 2; this rises to 97 per cent when the  {\sc magphys} parameters are added. There is no evidence that these dwarf galaxies can be separated into distinct types, despite our visual impression that this was the case. Instead they form a single population with a continuous distribution that can be parameterised by surface brightness or size (as in Figure~\ref{means}). {\it This implies that the different morphological classifications given to low luminosity star-forming galaxies in literature only reflect variations in surface brightness and do not correspond to any intrinsic physical differences between them}.  
 
\subsection{Limitations of the analysis and future prospects}
\label{limit}

 The \g II data analysed here suggests that the dwarf galaxies ($-18\!\leq\!M_r\!\leq\!-15$) in 
 the local Universe (8-87\,Mpc distance) form a unimodal population in the parameter space 
 spanned by $M_Z$, \re, $n$, \mue, $\mu_0$, $(g-i)$, $NUV-r$, $M^*$, $M_{dust}$, SFR and sSFR.
 This is most likely due to the intrinsic scatter resulting from stochastic evolution within different morphological classes of galaxies.
 Other factors may however also have affected the conclusions drawn here, and we discuss them in brief below.
 \begin{itemize}
 \item {\it Wavelength dependency:} The light profile of galaxies and the structural parameters derived from it are extremely
 wavelength-dependant. As shown by \citet{kelvin12} for a sample of more than $138,000$ galaxies in \g I, the S\'ersic index for 
 galaxies does not vary significantly with bandpass but minor variations are expected in individual galaxies because different
 optical wavebands trace different stellar components. On the other hand, the effective radius is a strong function of waveband in which
 the stellar light is modelled, such that the size of galaxy increases with wavelength \citep[][their figure 22]{kelvin12}. \citet{zhang12}
 reached similar conclusions when modelling ultraviolet to infrared data for the nearby dwarf irregular galaxies. 
 
 Although we employ the high-resolution {\it Viking} data in the $Z$-band for low surface brightness, low-luminosity galaxies here,
 the trends seen for the SDSS data used by \citet{kelvin12} still apply.  
 \item {\it Imaging data:} In this work we made use of the SDSS (Data Release 7) imaging to visually classify our sample into
 different classes. The SDSS images are taken with an exposure time of $\lesssim54$ seconds per field, and have an
 angular resolution of $1.2^{\prime\prime}$. Given that the bulk
 of our sample comprises low-luminosity dwarfs, the low-resolution imaging is a major limiting factor in the visual classification,
 and may have significantly contributed to the scatter within individual classes. This may also have caused morphologically similar
 classes such as the irregulars and the LSBs, or irregulars and the blue spheroids to overlap artificially in the parameter spaces explored
 in \S\ref{photo} and \S\ref{sf}. However, at this point we are unable to quantify this effect.
 
 The morphological classification of the sample presented here is $80\%$ reliable, meaning that among the visual classification by
 Mahajan, Driver and Drinkwater, and repetitive classification by Mahajan, the visual class assigned to galaxies agreed $\sim 80\%$
 of the times. The disagreement is mostly caused in the case of low surface brightness galaxies, and small, blue spheroids (the
 latter are as likely to be classified as irregulars). 
 
 The light profiles for our galaxy sample are modelled in the {\it Viking} $Z$-band data to automatically quantify structural parameters
 such as the S\'ersic index and effective radius using {\sc sigma}. The {\it Viking} data 
 are $\sim 2$ mag deeper and have a twofold improvement in angular resolution compared to the SDSS. 
 It is the best available dataset in the near infrared waveband for such a spectroscopically complete sample, yet still 
 shallow for dim or low-luminosity galaxies. Deeper datasets with higher resolution and depth are thus required for
 resolving the issues raised in this work due to measurement uncertainties, and confirm the observed trends in the structural
 and star formation properties of galaxies. 
     
 \item {\it Single component S\'ersic fit:} While a single component S\'ersic profile fits the giant elliptical galaxies 
 \citep[][and references therein]{tal11} and spirals \citep{pohlen06}  successfully, little is known of its behaviour at the faint end
  of the luminosity function.
 Recently though, \citet{herrmann13} have fitted single, double and triple exponential profiles to the stellar disk of 141 nearby dwarf galaxies
 in multi-wavelength data from ultraviolet to mid infrared. They find that blue compact dwarfs are over-represented by profiles
 where the light falls off less steeply, and Magellanic-like spirals by profiles where the light falls more steeply after the break in the
 first exponential, relative to dwarf irregulars. This observation is in agreement with the results presented by \citet{pohlen06},
 who showed  that $>90\%$ of their sample deviated from a classical $n=1$ exponential profile, and is instead better
 represented by a broken exponential profile.
 
 \citet{caon93} showed the correlation between S\'ersic index $n$ and (model-independent) size of early-type galaxies does not
 result from the parameter coupling in the S\'ersic model. This implies that if a model fails to capture the range of stellar distribution in
 a galaxy, it will result in over- or under-estimated magnitude, surface brightnesses and size as a function of the stellar concentration
 ($n$), and consequently $M^*$ \citep[see section 2.1 of the excellent review by][]{graham13}. 
 
 As we have shown in Figure~\ref{r-n} (also see Table~\ref{fit-stats}), while most bulge-dominated galaxies (blue spheroids and lenticulars) follow the correlation shown by \citet{caon93} for early-type and S0 cluster galaxies, spirals and other
 non-spheroidal dwarf galaxies do not.
 Table~\ref{fit-stats} suggests that almost half of the luminous spirals in our sample could be better fit by additional components to fit their
 light profile, especially in the centre. Some dwarf galaxies may also benefit by the inclusion of a second component. However, given the
 complex geometry of the low-luminosity irregular galaxies it is hard to believe that measurement uncertainties could be greatly improved
 by multiple component fits. Moreover, the measurement uncertainties obtained from these data are much smaller than the intrinsic
 scatter within each visually identifiable class. Hence, it is reasonable to assume that better models for the light profile may further tighten
 the correlations seen for the luminous galaxies. 
  \end{itemize}

 \section{summary}
 \label{summary}
 
 We have shown that morphologically distinct star-forming dwarf galaxies are not distinguishable in the parameter
 space comprising \mue, $\mu_0$, $n$, \re, SFR, sSFR, $M^*$, $M_{dust}$, ($NUV-r$) and ($g-i$). The (d)elliptical galaxies
 remain indistinguishable from the other dwarf classes in structural parameters, but their low sSFR and SFR makes them
 easily identifiable in star formation properties.
 In various 2D-parameter spaces formed by structural parameters, morphologically distinct dwarf galaxies occupy overlapping, yet different
 regions of the parameter space. However, all except the low-mass ellipticals also show similar star formation and dust properties. 
 The more luminous galaxies on the other hand, clearly separate into star-forming spirals and passively evolving lenticulars, respectively.
 For the ensemble of galaxies in our sample, the ``k-means" algorithm prefers a bimodal distribution
 independent of the number of parameters used to partition the data.

 We have shown for the first time the distribution of morphologically distinct dwarf galaxies in environments outside rich clusters
 in $M-\mu_0$ (\mue).
 Our analysis shows that although morphologically distinct galaxies occupy different regions of these spaces, in the dwarf regime
 the sub-populations overlap extensively. The giants on the other hand split into two as mentioned above.
   
 We showed that the model-independent correlation between stellar concentration, parametrized by the S\'ersic index $n$ and size (\re),
 shown to exist for elliptical and S0 cluster galaxies, is also followed by the blue spheroids and passively evolving ellipticals. The non-spheroidal galaxies on the other hand do not show such a correlation.
 
 The SFR, sSFR, $M_{dust}$ and $M^*$ derived by fitting the spectral energy distribution of all galaxies using {\sc magphys} qualitatively
 suggest that dwarf galaxies in these environments may evolve into red, passive low-mass ellipticals, while the
 luminous spiral galaxies turn into lenticulars. This hypothesis is based on the assumption that galaxies can not acquire much dust after
 their star formation turns off. A detailed analysis of the neutral hydrogen content of galaxies in these environments
 is required to confirm this speculation. 

 To conclude, we showed that galaxies across a wide range in magnitudes have statistically distinct, yet overlapping distributions
 in all their structural parameters. In the low-luminosity regime, the star-forming dwarf galaxies: LSBs, irregulars and blue spheroids
 also overlap in their star formation and dust properties, while the ellipticals have distinguishably low sSFR relative to other galaxies
 of similar luminosity. The giant galaxies
 on the other hand show clear separation such that the spirals are characterised by high SFR, sSFR and $M_{dust}$, while vice-versa
 is true for the lenticulars. 
 Hence, ``classification" (visual or otherwise) of galaxies forming a continuum in one, two, or multiple parameter
 space into discrete categories although useful in understanding the evolution mechanism(s) to first order, must be undertaken
 with caution, and consequent implications from such a classification must be used with caution.

\section{Acknowledgements}
 GAMA is a joint European-Australasian project based around a spectroscopic campaign using the
 Anglo-Australian Telescope. The GAMA input catalogue is based on data taken from the Sloan Digital Sky
 Survey and the UKIRT Infrared Deep Sky Survey. Complementary imaging of the GAMA regions is being
 obtained by a number of independent survey programs including GALEX MIS, VST KiDS, VISTA VIKING, 
 WISE, Herschel-ATLAS, GMRT and ASKAP providing UV to radio coverage. GAMA is funded by the STFC
 (UK), the ARC (Australia), the AAO, and the participating institutions.
 The GAMA website is http://www.gama-survey.org/. 
 
 This project was partly funded by the UWA-UQ Bilateral Research Collaboration Awards (University of Queensland,
 2013; CI: Mahajan).

 We thank Michael Brown, Alister Graham, Maritza Lara-Lopez, Trevor Ponman and Stephen Wilkins for
 their comments and suggestions on an early draft of this paper. We also thank the anonymous referee for
 their constructive criticism which greatly helped in clarifying the presentation of this work.


\appendix
\section*{Appendix}
\renewcommand{\thesubsection}{\Alph{subsection}}

\subsection[]{Kolmogorov-Smirnov Statistics}
\label{ks:table}

In this appendix, we present tables of probabilities from the Kolmogorov-Smirnov static, showing that the physical and star formation properties
of various visual morphological classes arise from the same parent distributions. 

\setcounter{table}{0}
\renewcommand{\thetable}{A\arabic{table}}

\begin{table}
\caption{K-S probabilities for log $M^*$.}
\begin{tabular}{|c|c|c|c|c|c|c|}
\hline 
          &  E &  Sp  &   Irr  &   BSph   &   LSB    &   Len  \\ \hline
      E &  1  & 1.98E-01 & \bf{1.17E-06} & \bf{3.99E-05} & \bf{9.63E-08} & \bf{3.86E-08} \\
      Sp &  & 1 & \bf{6.21E-14} & \bf{1.34E-08} & \bf{1.55E-13} &  \bf{3.41E-06}  \\
      Irr &   &   &  1 & 1.05E-01 &  \bf{3.61E-03} & \bf{1.66E-20}  \\
      BSph &   &   &  & 1 &  \bf{3.54E-02} & \bf{2.64E-16}  \\
      LSB & &  &  &  & 1 &  \bf{6.07E-18}  \\
     Len &   &   &  &   &  &  1 \\        
\hline
\end{tabular}
\end{table}

\begin{table}
\caption{K-S probabilities for SFR.}
\begin{tabular}{|c|c|c|c|c|c|c|}
\hline 
          &  E &  Sp  &   Irr  &   BSph   &   LSB    &   Len  \\ \hline
 E   &  1 &   \bf{4.25E-08} &   \bf{3.27E-03} &   \bf{6.02E-03} &   \bf{2.61E-02} &  5.84E-01  \\
 Sp &   & 1 &  \bf{1.26E-16}  &  \bf{2.76E-15} &  \bf{1.11E-17} &  \bf{8.92E-09}  \\
 Irr & &  & 1 &  4.66E-01 &  \bf{9.39E-04} &   \bf{8.00E-03}  \\
 BSph & &   &  &  1 &   \bf{2.44E-02} &   \bf{1.99E-02}  \\
 LSB &   &  &  &  & 1  & 7.98E-02  \\
 Len &   &   &  &   &  &  1 \\        
\hline
\end{tabular}
\end{table}

\begin{table}
\caption{K-S probabilities for SFR/$M^*$.}
\begin{tabular}{|c|c|c|c|c|c|c|}
\hline 
          &  E &  Sp  &   Irr  &   BSph   &   LSB    &   Len  \\ \hline
 E &  1 & \bf{8.51E-06} &  \bf{2.41E-12} &  \bf{4.78E-10} &  \bf{4.20E-09} &  \bf{1.45E-05}  \\
 Sp &    & 1 & \bf{2.23E-03} &  \bf{8.16E-03} &  \bf{3.08E-02} &  \bf{7.10E-10}  \\
 Irr &   &   & 1  & 2.41E-01 &  7.24E-01  & \bf{2.26E-17}  \\
 BSph &  &  &  & 1 &  2.27E-01 &  \bf{6.59E-14} \\
 LSB &  & &  &   & 1 &  \bf{5.59E-14}  \\
 Len &   &   &  &   &  &  1 \\        
\hline
\end{tabular}
\end{table}

\begin{table}
\caption{K-S probabilities for $M_{dust}$.}
\begin{tabular}{|c|c|c|c|c|c|c|}
\hline 
          &  E &  Sp  &   Irr  &   BSph   &   LSB    &   Len  \\ \hline
 E &  1 & \bf{3.35E-08} &  3.58E-01 &  6.73E-02 &  \bf{3.70E-02} &  \bf{1.70E-04}  \\
 Sp &   & 1 & \bf{3.63E-21} &  \bf{1.04E-17} &  \bf{7.35E-20} &  \bf{2.87E-02}   \\
 Irr &   &  & 1  & 1.51E-01 &  1.08E-01 &  \bf{1.05E-09}  \\
 BSph &   & &  & 1 & 6.76E-01 &  \bf{3.12E-09}  \\
 LSB &  &  &  &  & 1 &  \bf{3.85E-10}  \\         
 Len &   &   &  &   &  &  1 \\        
\hline
\end{tabular}
\end{table}

\begin{table}
\caption{K-S probabilities for \re.}
\begin{tabular}{|c|c|c|c|c|c|c|}
\hline 
          &  E &  Sp  &   Irr  &   BSph   &   LSB    &   Len  \\ \hline
 E &  1 & \bf{1.14E-07} &  6.66E-01 &  5.66E-02 &  4.61E-01 &  \bf{5.09E-03}  \\
 Sp &   & 1 & \bf{3.27E-15} &  \bf{3.40E-16} &  \bf{5.86E-12} &  \bf{1.51E-04}  \\
 Irr &  &  & 1 &  \bf{2.84E-06} &  5.19E-01 &  \bf{3.61E-04}  \\
 BSph &  &  &  & 1 &  \bf{5.79E-06} &  \bf{5.33E-10} \\
 LSB &  &  &  &  & 1 & \bf{2.21E-03}  \\          
 Len &   &   &  &   &  &  1 \\        
\hline
\end{tabular}
\end{table}

\begin{table}
\caption{K-S probabilities for $n$.}
\begin{tabular}{|c|c|c|c|c|c|c|}
\hline 
          &  E &  Sp  &   Irr  &   BSph   &   LSB    &   Len  \\ \hline
  E &  1 & 1.88E-01 &  \bf{9.83E-03} &  9.37E-01 &  \bf{2.61E-02} &  \bf{1.28E-04}  \\
 Sp &  & 1 &  5.27E-01 &  \bf{5.23E-03} &  4.55E-01 &  \bf{7.10E-10}  \\
 Irr &  &  & 1 & \bf{1.80E-06} &  4.32E-01 &  \bf{1.23E-13}  \\
 BSph &  &  &  & 1 & \bf{2.43E-04} &  \bf{2.33E-08}  \\
 LSB &  &  &  &  & 1  & \bf{1.22E-10} \\         
 Len &   &   &  &   &  &  1 \\        
\hline
\end{tabular}
\end{table}

\begin{table}
\caption{K-S probabilities for $\mu_0$.}
\begin{tabular}{|c|c|c|c|c|c|c|}
\hline 
          &  E &  Sp  &   Irr  &   BSph   &   LSB    &   Len  \\ \hline
  E &  1 & 7.67E-01 &  \bf{5.90E-03} & 5.05E-01 & \bf{8.93E-09} &  \bf{2.46E-06}  \\
 Sp &  & 1 & \bf{1.05E-07} &  4.54E-01 &  \bf{9.01E-15} &  \bf{8.92E-10} \\ 
 Irr & & & 1 & \bf{7.81E-07} &  \bf{8.57E-11} &  \bf{1.14E-17} \\
 BSph &  &  &  & 1 & \bf{3.66E-18} &  \bf{2.98E-14}  \\
 LSB &   &  &  &  & 1 & \bf{1.44E-14}  \\        
 Len &   &   &  &   &  &  1 \\        
\hline
\end{tabular}
\end{table}

\begin{table}
\caption{K-S probabilities for \mue.}
\begin{tabular}{|c|c|c|c|c|c|c|}
\hline 
          &  E &  Sp  &   Irr  &   BSph   &   LSB    &   Len  \\ \hline
  E &  1 & 1.06E-01 & \bf{2.78E-02} &  5.49E-01 &  \bf{1.77E-10} & \bf{1.36E-05}  \\
 Sp &  & 1 & \bf{4.72E-09} &  \bf{3.12E-03} &  \bf{3.65E-20} &  \bf{1.47E-05}  \\
 Irr &  &  & 1 & \bf{2.76E-05} &  \bf{2.81E-18} & \bf{1.67E-14}  \\
 BSph &   &  &   &  1 & \bf{1.60E-23} &  \bf{2.90E-12}  \\
 LSB & &  &  &  & 1 & \bf{5.69E-16} \\         
 Len &   &   &  &   &  &  1 \\        
\hline
\end{tabular}
\end{table}

\subsection{Preferred number of clusters with k-means clustering algorithm}
\label{nclusters}

In this Appendix we illustrate the Monte Carlo approach we used to estimate uncertainties in the statistics used to select the best number of clusters from the k-means analysis. As noted above (Section~\ref{s:nclus}), we used the \citet{calinski74} variance ratio as a figure-of-merit. We used the NbClust package \citep{charrad13} for all the calculations, but this did not provide uncertainties for the calculated variance ratios. We applied a simple Monte-Carlo approach to estimate the variance ratio uncertainties by repeating each calculation 30 times using a different random seed for the k-means partition algorithm each time. We used the mean value of the 30 calculations as the final value of the statistic and the standard deviation of the 30 values to estimate an uncertainty (the standard error of the mean) which we show using the red symbols in Figure~\ref{varplot} for each of the three partitions shown in Figure~\ref{kmeans}. As the standard errors on each point are very small we also show the standard deviations (in grey) for comparison. We also show the results of a single calculation, showing that quite strong systematic trends can be present (e.g.\ for 6-10 clusters in the first panel): this is an additional reason to take the mean of several calculations.

\setcounter{figure}{0}
\renewcommand{\thefigure}{B\arabic{figure}}

\begin{figure}
\centering{
{\rotatebox{270}{\epsfig{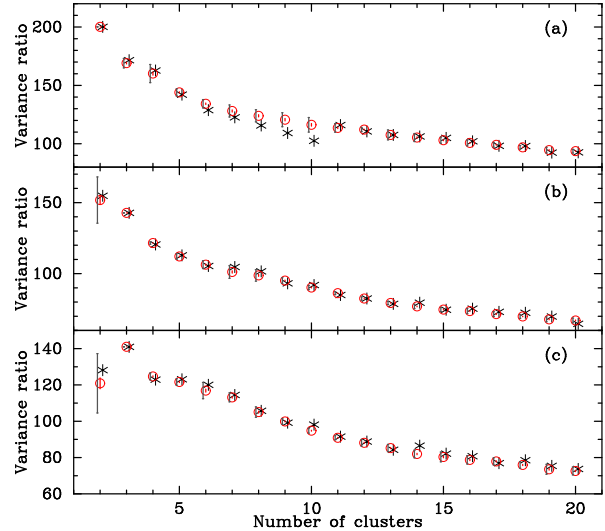}}}}
\caption{The k-means variance ratio as a function of the number of clusters analysed. The three plots are for the same parameter sets as in Figure~\ref{kmeans}: (a) Structural (b) Structural+SF 1, and (c) Structural+SF 3, respectively as discussed above and listed in Table~\ref{clusterlist}. In each plot the black stars represent a single k-means calculation using a random seed of 12341. The red circles show the mean of 30 calculations using different random seeds. The red error bars give the standard  error of the mean and the grey error bars show the 1-standard deviation range of the values. We select the best number of clusters by identifying a local maximum in the variance ratio: only in plot (c) is a 3-cluster partition preferred over the 2-cluster case.}
\label{varplot}
\end{figure}

In panels (a) and (c) of Figure~\ref{varplot} there is a clear local maximum of the variance ratio (at 2,3 clusters respectively) that is separated by more than 3 standard errors (the smaller error bars) from the adjacent points. In panel (b) the first point is barely three standard errors higher than the second, but we take the preferred number of clusters as two using our rule to take the smaller number of clusters in the case of a double peak. 


\label{lastpage}


\begin{thebibliography}{99}

\bibitem[\protect\citeauthoryear{Baldry, Glazebrook, \& Driver}{2008}]{baldry08} Baldry I.~K., Glazebrook K.,
 Driver S.~P., 2008, MNRAS, 388, 945 
\bibitem[\protect\citeauthoryear{Baldry et al.}{2010}]{baldry10} Baldry I.~K., et al., 2010, MNRAS, 404, 86
\bibitem[\protect\citeauthoryear{Baldry et al.}{2012}]{baldry12} Baldry I.~K., et al., 2012, MNRAS, 421, 621 
\bibitem[\protect\citeauthoryear{Bertin \& Arnouts}{1996}]{bertin96} Bertin E., Arnouts S., 1996, A\&AS, 117, 393 
\bibitem[\protect\citeauthoryear{Bertin}{2011}]{bertin11} Bertin, E., 2011, Astronomical 
Data Analysis Software and Systems XX, 442, 435 
\bibitem[\protect\citeauthoryear{Binggeli, Sandage, 
\& Tarenghi}{1984}]{binggeli84} Binggeli B., Sandage A., Tarenghi M., 1984, AJ, 89, 64 
\bibitem[\protect\citeauthoryear{Binggeli}{1994}]{binggeli94} Binggeli B., 1994, ESOC, 49, 13
\bibitem[\protect\citeauthoryear{Blanton et al.}{2005}]{blanton05} Blanton M.~R., Lupton R.~H., Schlegel 
D.~J., Strauss M.~A., Brinkmann J., Fukugita M., Loveday J., 2005, ApJ, 631, 208 
\bibitem[\protect\citeauthoryear{Boselli et al.}{2008}]{boselli08} Boselli A., Boissier S., Cortese L.,
 Gavazzi G., 2008, A\&A, 489, 1015 
\bibitem[\protect\citeauthoryear{Bothun et al.}{1987}]{bothun87} Bothun G.~D., Impey C.~D., Malin D.~F.,
 Mould J.~R., 1987, AJ, 94, 23 
\bibitem[\protect\citeauthoryear{Bourne et al.}{2012}]{bourne12} Bourne N., et al., 2012, MNRAS, 421, 3027  
\bibitem[\protect\citeauthoryear{Boyce \& Phillipps}{1995}]{boyce95} Boyce P.~J., Phillipps S., 1995, A\&A, 296, 26 
\bibitem[\protect\citeauthoryear{Brosche}{1973}]{brosche73} Brosche P., 1973, A\&A, 23, 259
\bibitem[\protect\citeauthoryear{Brough et al.}{2013}]{brough13} Brough S., et al., 2013, MNRAS, 435, 2903 

\bibitem[\protect\citeauthoryear{Caldwell}{1983}]{caldwell83} Caldwell N., 1983, AJ, 88, 804 
\bibitem[\protect\citeauthoryear{Calinski \& Harabasz}{1974}]{calinski74} Calinski T., Harabasz J., 1974,
Communications in Statistics - Theory and Methods, 3(1), 1-27. doi:10.1080/03610927408827101. 
\bibitem[\protect\citeauthoryear{Caon, Capaccioli, \& D'Onofrio}{1993}]{caon93} Caon N., Capaccioli M.,
 D'Onofrio M., 1993, MNRAS, 265, 1013 
\bibitem[\protect\citeauthoryear{Caon \& Einasto}{1995}]{caon95} Caon N., Einasto M., 1995, MNRAS, 273, 913 
\bibitem[\protect\citeauthoryear{Cortese et al.}{2011}]{cortese11} Cortese L., Catinella B., Boissier S., 
Boselli A., Heinis S., 2011, MNRAS, 415, 1797
\bibitem[\protect\citeauthoryear{Charrad et al.}{2012}]{charrad13} Charrad M., Ghazzali N., Boiteau V., Niknafs A., 2012, 
 UseR! 2012, CGB12a

\bibitem[\protect\citeauthoryear{da Cunha, Charlot, \& Elbaz}{2008}]{dacunha08} da Cunha E., Charlot S., Elbaz D.,
 2008, MNRAS, 388, 1595 
\bibitem[\protect\citeauthoryear{Dressler}{1980}]{dressler80} Dressler A., 1980, ApJ, 236, 351 
 \bibitem[\protect\citeauthoryear{Driver et al.}{2011}]{driver11} Driver S.~P., et al., 2011, MNRAS, 413, 971
 \bibitem[\protect\citeauthoryear{Driver et al.}{2012}]{driver12} Driver S.~P., et al., 2012, MNRAS, 427, 3244 

  \bibitem[\protect\citeauthoryear{Fujita}{1998}]{fujita98} Fujita, Y., 1998, ApJ, 509, 587
 
 \bibitem[\protect\citeauthoryear{Graham}{2013}]{graham13} Graham A.~W., 2013, pss6.book, 91 
 \bibitem[\protect\citeauthoryear{Graham \& Guzm{\'a}n}{2003}]{graham03} Graham A.~W.,
 Guzm{\'a}n R., 2003, AJ, 125, 2936 
 \bibitem[\protect\citeauthoryear{Graham \& Driver}{2005}]{graham05} Graham A.~W.,
 Driver S.~P., 2005, PASA, 22, 118 

\bibitem[\protect\citeauthoryear{Herrmann, Hunter, \& Elmegreen}{2013}]{herrmann13} Herrmann K.~A.,
 Hunter D.~A., Elmegreen B.~G. 2013, AJ, 146, 104 
 \bibitem[\protect\citeauthoryear{Hill et al.}{2011}]{hill11} Hill D.~T., et al., 2011, MNRAS, 412, 765 
\bibitem[\protect\citeauthoryear{Hubble}{1926}]{hubble26} Hubble E.~P., 1926, ApJ, 64, 321 
\bibitem[\protect\citeauthoryear{Hunter, Elmegreen, \& Ludka}{2010}]{hunter10} Hunter D.~A., Elmegreen B.~G.,
 Ludka B.~C., 2010, AJ, 139, 447 

\bibitem[\protect\citeauthoryear{Janz et al.}{2013}]{janz13} Janz J., et al., 2013, arXiv, arXiv:1308.6496 
\bibitem[\protect\citeauthoryear{Janz et al.}{2012}]{janz12} Janz J., et al., 2012, ApJ, 745, L24 
\bibitem[\protect\citeauthoryear{Jenkins et al.}{2007}]{jenkins07} Jenkins L.~P., Hornschemeier A.~E., 
 Mobasher B., Alexander D.~M., Bauer F.~E., 2007, ApJ, 666, 846 

 \bibitem[\protect\citeauthoryear{Lake, Katz, \& Moore}{1998}]{lake98} Lake G., Katz N., Moore B.,
  1998, ApJ, 495, 152
 \bibitem[\protect\citeauthoryear{Lee et al.}{2013}]{lee13} Lee B., et al., 2013, ApJ, 774, 47 
 
\bibitem[\protect\citeauthoryear{Oh \& Lin}{2000}]{oh2000} Oh K.~S., Lin D.~N.~C., 2000, ApJ, 543, 620 

 \bibitem[\protect\citeauthoryear{Kaviraj et al.}{2007}]{kaviraj07} Kaviraj S., et al., 2007, ApJS, 173, 619 
 \bibitem[\protect\citeauthoryear{Kelvin et al.}{2012}]{kelvin12}  Kelvin L.~S., et al., 2012, MNRAS, 421, 1007 
\bibitem[\protect\citeauthoryear{Kodaira, Okamura, \& Watanabe}{1983}]{kodaira83}
 Kodaira K., Okamura S., Watanabe M., 1983, ApJ, 274, L49 
\bibitem[\protect\citeauthoryear{Kodaira}{1989}]{kodaira89} Kodaira K., 1989, ApJ, 342, 122
\bibitem[\protect\citeauthoryear{Kormendy}{1985}]{kormendy85} Kormendy J., 1985, ApJ, 295, 73
 \bibitem[\protect\citeauthoryear{Kormendy \& Bender}{2012}]{kormendy12} Kormendy J.,
  Bender R., 2012, ApJS, 198, 2

\bibitem[\protect\citeauthoryear{Loveday et al.}{2012}]{loveday12} Loveday J., et al., 2012, MNRAS, 420, 1239 

 \bibitem[\protect\citeauthoryear{MacQueen}{1967}]{macqueen67} Macqueen J., ``Some methods for classification and analysis of
  multivariate observations", Proc. Fifth Berkeley Symp. on Math. Statist. and Prob., Vol. 1 (Univ. of Calif. Press, 1967), 281, 297
 \bibitem[\protect\citeauthoryear{Mahajan, Raychaudhury, \& Pimbblet}{2012}]{mahajan12} Mahajan S.,
  Raychaudhury S., Pimbblet K.~A., 2012, MNRAS, 427, 1252 
 \bibitem[\protect\citeauthoryear{McConnachie}{2012}]{mcconnachie12} McConnachie A.~W., 2012, AJ, 144, 4 

\bibitem[\protect\citeauthoryear{Peng et al.}{2010}]{peng10} Peng C.~Y., Ho L.~C., Impey C.~D.,
 Rix H.-W., 2010, AJ, 139, 2097 
\bibitem[\protect\citeauthoryear{Pohlen \& Trujillo}{2006}]{pohlen06} Pohlen M., Trujillo I., 2006, A\&A, 454, 759 
\bibitem[\protect\citeauthoryear{Popesso et al.}{2006}]{popesso06} Popesso P., Biviano A., B{\"o}hringer H.,
 Romaniello M., 2006, A\&A, 445, 29  
\bibitem[\protect\citeauthoryear{Prugniel \& Simien}{1997}]{prugniel97} Prugniel P., Simien F.,
 1997, A\&A, 321, 111 

\bibitem[\protect\citeauthoryear{Reynolds}{1920}]{reynolds20} Reynolds J.~H., 1920, MNRAS, 80, 746 
\bibitem[\protect\citeauthoryear{Robotham et al.}{2011}]{robotham11} Robotham A.~S.~G., et al., 2011, MNRAS,
 416, 2640 

\bibitem[\protect\citeauthoryear{Schawinski et al.}{2007}]{schawinski07} Schawinski K.,
 et al., 2007, ApJS, 173, 512 
 \bibitem[\protect\citeauthoryear{Sharina et al.}{2008}]{sharina08} Sharina M.~E., et al., 2008, MNRAS, 384, 1544 
 \bibitem[\protect\citeauthoryear{Strader et al.}{2013}]{strader13} Strader J., et al., 2013, ApJ, 775, L6 
 
 \bibitem[\protect\citeauthoryear{Tal \& van Dokkum}{2011}]{tal11} Tal T., van Dokkum P.~G., 2011, ApJ, 731, 89
 \bibitem[\protect\citeauthoryear{Taylor et al.}{2011}]{taylor11} Taylor E.~N., et al., 2011, MNRAS, 418, 1587  
 \bibitem[\protect\citeauthoryear{Thilker et al.}{2007}]{thilker07} Thilker D.~A., et al., 2007, ApJS, 173, 538 
 \bibitem[\protect\citeauthoryear{Thomas, Drinkwater, \& Evstigneeva}{2008}]{thomas08} Thomas P.~A.,
 Drinkwater M.~J., Evstigneeva E., 2008, MNRAS, 389, 102 
 \bibitem[\protect\citeauthoryear{Toloba et al.}{2011}]{toloba11} Toloba E., Boselli A., Cenarro A.~J., Peletier R.~F.,
  Gorgas J., Gil de Paz A., Mu{\~n}oz-Mateos J.~C., 2011, A\&A, 526, A114 

\bibitem[\protect\citeauthoryear{Wright et al.}{2010}]{wright10} Wright E.~L., et al., 2010, AJ, 140, 1868 

 \bibitem[\protect\citeauthoryear{Yamanoi et al.}{2012}]{yamanoi12} Yamanoi H., et al., 2012, AJ, 144, 40
 \bibitem[\protect\citeauthoryear{Young \& Currie}{1995}]{young95} Young C.~K., Currie M.~J., 1995,
 MNRAS, 273, 1141 

\bibitem[\protect\citeauthoryear{Zhang et al.}{2012}]{zhang12} Zhang H.-X., Hunter D.~A., Elmegreen B.~G., Gao Y.,
 Schruba A., 2012, AJ, 143, 47 

 \end{thebibliography}
\end{document}